\documentclass[preprint,journal]{vgtc}       

\ifpdf
  \pdfoutput=1\relax                   
  \usepackage{graphicx}                
  \DeclareGraphicsExtensions{.pdf,.png,.jpg,.jpeg} 
\else
  \usepackage{graphicx}                
  \DeclareGraphicsExtensions{.eps}     
\fi%

\graphicspath{{figures/}{pictures/}{images/}{./}} 

\usepackage{microtype}                 
\PassOptionsToPackage{warn}{textcomp}  
\usepackage{textcomp}                  
\usepackage{mathptmx}                  
\usepackage{times}                     
\usepackage{cite}                      
\usepackage{tabu}                      
\usepackage{booktabs}                  
\usepackage{wrapfig}
\usepackage{enumitem}

\usepackage[final]{changes}

\ieeedoi{10.1109/TVCG.2020.3030435}

\onlineid{0}

\vgtccategory{Research}
\vgtcpapertype{please specify}

\title{Data Visceralization: Enabling Deeper Understanding of Data Using Virtual Reality}

\author{Benjamin Lee, Dave Brown, Bongshin Lee, Christophe Hurter, Steven Drucker, and Tim Dwyer}
\authorfooter{
\item
 Benjamin Lee and Tim Dwyer are with Monash University. Email: \{benjamin.lee1,tim.dwyer\}@monash.edu.
\item
 Dave Brown, Bongshin Lee, and Steven Drucker are with Microsoft Research. Email: \{dave.brown,bongshin,sdrucker\}@microsoft.com.
\item
 Christophe Hurter is with ENAC, French Civil Aviation University. Email: christophe.hurter@enac.fr.
}

\shortauthortitle{Lee \MakeLowercase{\textit{et al.}}: Data Visceralization: Enabling Deeper Understanding of Data Using Virtual Reality}

\abstract{
A fundamental part of data visualization is transforming data to map abstract information onto visual attributes. While this abstraction is a powerful basis for data visualization, the connection between the representation and the original underlying data (i.e., what the quantities and measurements actually correspond with in reality) can be lost. On the other hand, virtual reality (VR) is being increasingly used to represent real and abstract models as natural experiences to users. In this work, we explore the potential of using VR to help restore the basic understanding of units and measures that are often abstracted away in data visualization in an approach we call \textit{data visceralization}. 
By building VR prototypes as design probes, we identify key themes and factors for data visceralization. We do this first through a critical reflection by the authors, then by involving external participants.
We find that data visceralization is an engaging way of understanding the qualitative aspects of physical measures and their real-life form, which complements analytical and quantitative understanding commonly gained from data visualization.
However, data visceralization is most effective when there is a one-to-one mapping between data and representation, with transformations such as scaling affecting this understanding.
We conclude with a discussion of future directions for data visceralization.
} 


\CCScatlist{ 
 \CCScat{K.6.1}{Management of Computing and Information Systems}%
{Project and People Management}{Life Cycle};
 \CCScat{K.7.m}{The Computing Profession}{Miscellaneous}{Ethics}
}

\teaser{
  \centering
 \includegraphics[width=0.8\linewidth]{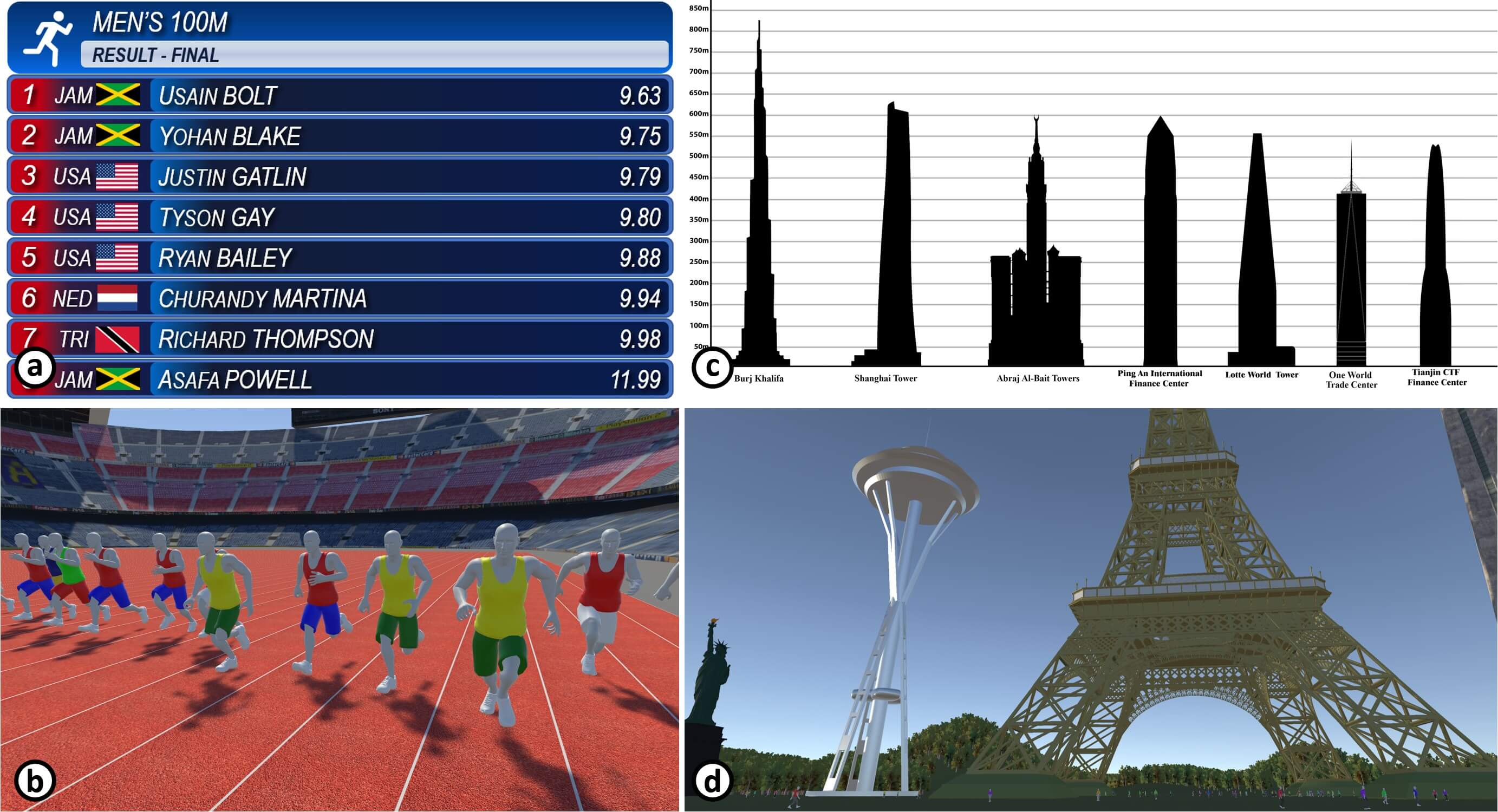}
  \caption{Prototypes of \textit{data visceralizations} in VR based on popular representations of physical measurements. (a) \replaced{Scorecard}{table} of results in seconds from Olympic Men's 100~m. (b) Data visceralization equivalent to experience one-to-one scale of Olympic sprint speeds. (c) Comparison diagram of tall skyscrapers \added{(\textcopyright{} Saggittarius A, CC BY-SA 4.0)}. (d) Data visceralization equivalent to experience and compare true one-to-one scale of select skyscrapers.}
	\label{fig:teaser}
}

\newcommand{\bpstart}[1]{\vspace{0.5mm} \noindent{\textbf{#1.}}}

\vgtcinsertpkg

\begin{document}


\firstsection{Introduction}

\maketitle

Communicating information using stories that employ data visualization has been explored extensively in recent years \cite{Riche:2018:DDS, Segel:2010:NVT}. 
A fundamental part of data visualization is processing and transforming raw data, ultimately mapping this abstracted information into attributes represented in a visualization \cite{Card:1999:RIV, Chi:1998:OIF}. This abstraction, while powerful and in many cases necessary, poses a limitation for data based on physical properties, where the process of measurement causes the connection between the visualization and the underlying `meaning' of the data to be lost (i.e., what the data truly represents in the real-world). While techniques in data-driven storytelling (e.g.,~\cite{stolper2018data}) can help establish context and resolve ambiguity in these cases, these techniques do little to help people truly \textit{understand} the underlying data itself. 
A common approach used to help improve comprehension of these measures is by using concrete scales \cite{Chevalier:2013:UCS}---the association of physical measurements and quantities with more familiar objects. However, this often relies on prior knowledge and requires cognitive effort to effectively envision the desired mental imagery.

\begin{figure}[t]
    \centering
    \includegraphics[width=0.95\linewidth]{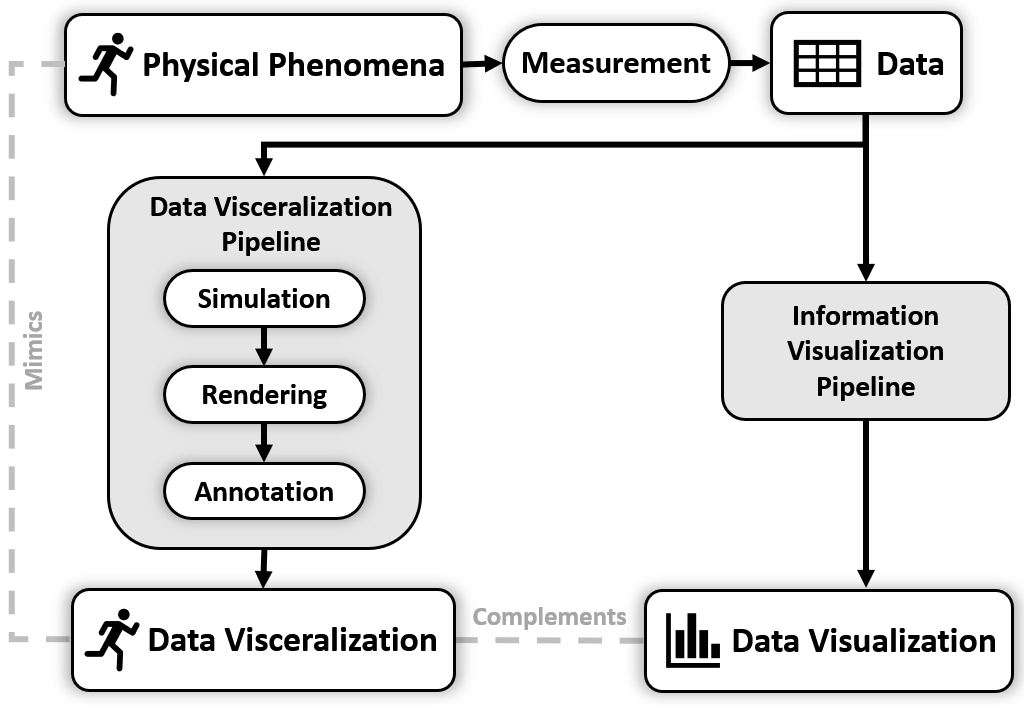}
    \caption{Our conceptual data visceralization pipeline in relation to the information visualization pipeline \cite{Card:1999:RIV}. Both run in parallel to each other, with data visceralization aimed at complementing data visualization.
    } 
    \label{fig:pipeline}
\end{figure}

To complement these approaches in data visualization and storytelling, we introduce \textit{data visceralization}, which we define as a data-driven experience which evokes visceral feelings within a user to facilitate intuitive understanding of physical measurements and quantities. By visceral, we mean a ``subjective sensation of being there in a scene depicted by a medium, usually virtual in nature'' \cite{Barfield:1995:VEA, Barfield:1995:PPW}. To illustrate this concept, consider the scenarios depicted in Fig.~\ref{fig:teaser}(a,c). Many of us are familiar with 100~m sprint times, but understanding the `ground truth'---how fast they are \textit{actually} running---is elusive. Similarly, images or diagrams showcasing tall buildings such as skyscrapers are common, but mentally envisioning what these actually look like without prior knowledge is challenging. In these scenarios, only by seeing the actual sprinter or building will achieve this truth. In a sense, seeing---or more generally, experiencing---is believing.
With virtual reality (VR) technology rapidly advancing and becoming more readily available, it offers an unprecedented opportunity for achieving these visceral experiences in a manner that is both cost effective and compelling. As depicted in Fig.~\ref{fig:teaser}(b,d), we can now simulate what it's like to have Olympic sprinters run right past you, or the feeling of being next to one of these skyscrapers.
From these experiences, a thorough understanding of these measures and quantities may be achieved, with data visualization and visceralization existing hand-in-hand to provide both quantitative (i.e., analytical reasoning) and qualitative (i.e., the ground truth) understanding of the data (Fig.~\ref{fig:pipeline}).

In this work, we explore this concept of data visceralization. We develop six VR prototypes as design probes based on existing data stories and visualizations specifically chosen to explore a range of different measures and phenomena.
We critically reflect on these design probes, identifying key themes and factors for data visceralization such as: the appropriate types of measures and quantities; the ranges of magnitudes of physical phenomena that are suitable; and the situations where they are effective or not.
We expand this reflection through sessions with external participants to gain feedback on the value and intricacies of data visceralization.
We conclude by discussing multiple aspects of data visceralizations, along with future work in the area.

In summary, the main contributions of this paper are:
\vspace{-2mm}
\begin{itemize}[leftmargin=*, noitemsep]
    \item The introduction of the novel concept of data visceralization, and its applications for understanding the data that underlies visualizations
    \item A set of prototype examples to demonstrate the concept and characterize the experiences
    \item An exploration into the factors and considerations behind data visceralization, through both a critical reflection by the authors and external feedback from participants
\end{itemize}

\section{Related work} 
\label{sec:related-work}
While our work focuses on virtual reality (VR), the concept of data visceralization can be applied much more broadly. There are other non-VR oriented methods of helping people understand a unit or measure such as scale models, immersive IMAX films, museum exhibits \cite{Mortenson:2011}, or first hand experiences. In this section, we discuss data visceralization in the context of other related fields.


\subsection{(Immersive) Data-Driven Storytelling}
Segel and Heer \cite{Segel:2010:NVT} coined the term \textit{narrative visualization} in 2010.
Since then, researchers have explored the design patterns of specific genres of narrative visualizations, such as Amini et al. \cite{Amini:2015:UDV:2702123.2702431} with data videos and Bach et al. \cite{Bach:2018:NDP} with data comics. In contrast, work by Stolper et al. \cite{stolper2018data} characterized the range of recently emerging techniques used in narrative visualization as a whole.
With the increased use of VR and augmented reality (AR) devices for the purposes of data visualization and analytics (known as \textit{immersive analytics} \cite{Marriott:2018:IA}) and for storytelling in general \cite{Bucher:2017:SVR}, it is feasible to begin considering how these devices can be used for \textit{immersive} data-driven storytelling \cite{Isenberg:2018:IVD,lee2018watches}.
Recent work by Ren et al. \cite{Ren:2018:XIC} has investigated the creation of immersive data stories, and work by Bastiras et al. \cite{Bastiras:2017:CVR} of their effectiveness, but these resort to using simple 2D images in a VR environment rather than taking full advantage of the device's capabilities.
A VR piece from the Wall Street Journal \cite{WallStreetJournal:2015:3DNasdaq} remains one of the few compelling examples of an immersive data story, using a time series of the NASDAQ's price/earnings ratio over 21 years as a roller coaster track which the reader then rides from one end to the other. The story particularly focuses on the sudden fall in the index as the dot-com bubble began to burst in 2000, having readers experience this metaphorical fall as a literal roller coaster drop in VR.
In this work, we examine the use of immersive environments to achieve similar effects of transforming data into visceral experiences, but in a complementary fashion to existing data stories. That is, we focus on aiding the understanding of the underlying data itself through the use of VR, while the narrative and visualizations of the story set the context, background, and messaging.


\subsection{Concrete Scales and Personalization}
The use of concrete scales is a popular technique used to aid in comprehending unfamiliar units and extreme magnitudes by comparison to more familiar objects. To formalize this technique, Chevalier et al. \cite{Chevalier:2013:UCS} collected and analyzed over 300 graphic compositions found online. They derived a taxonomy of object types and measure relations, and identified common strategies such as using analogies of pairwise comparison and juxtaposing multiples of smaller objects together. They discussed the need and challenges of choosing good units, a problem which Hullman et al. \cite{Hullman:2018:ICM} addressed by automatically generating different re-expression units which may be more familiar to the user. Concrete scales fundamentally assists in building mental models of scale. In contrast, we use data visceralization to directly represent the measure that is being conveyed to the user, effectively skipping this process altogether.


\subsection{Data Physicalization} \label{ssc:data-phys}
As compared to information visualization which maps data into visual marks and variables \cite{Card:1999:RIV, Chi:1998:OIF}, data physicalization explores how data can be encoded and presented in tangible, physical artifacts through its geometric and/or material properties \cite{Jansen:2013:EEP, Jansen:2015:OCD}. Although users can directly experience  data in a unique physicalized manner, it still fundamentally transforms and remaps abstract data into tangible experiences, often in equivalents of common visualization types \cite{Jansen:2013:EEP, Taher:2017:IUD, Dragicevic:2019:DataPhysList}. While this tangibility may provide benefits for memory or engagement, the focus is on some higher level conceptual data rather than the physical property or measure itself. Data physicalization could be well suited for creating visceral experiences if attributes are represented without transformation, as any physical phenomena can theoretically be fabricated and subsequently experienced. However, this would be resource intensive and heavily dependent on advances in technology. Indeed, many museum exhibits construct one-to-one mappings of data phenomena so that people can understand the underlying data in representation, but such exhibits are expensive and also can only be experienced by visiting them. Therefore technologies such as VR are well suited for visceralization, overcoming barriers such as fabrication cost and physical space restrictions through use of virtual locomotion techniques \cite{Laviola:2017:3DU}.

\subsection{Immersion and Presence in VR}
VR and immersive technologies as a whole have been available for many decades, and have been extensively studied for their impact on human perception. Core to these technologies are the notions of immersion and presence, where immersion is a characteristic of the technology that enables a vivid illusion of reality to the user, and presence is the state of consciousness of \textit{being} in the virtual environment \cite{Slater:1997:FIV, Slater:1999:MPR}. Presence is a form of visceral communication that is primal but also difficult to describe \cite{Jerald:2015:VBH}. This notion of presence and viscerally believing that the virtual world is real is what we aim to leverage with visceralization. VR devices have proven effective enough in doing this that they have been used in applications such as psychological treatment \cite{Valmaggia:2016:VRP}, journalism \cite{delaPena:2010:IJI}, and military training \cite{Lele:2013:VRM}. Moreover, VR and AR has been used in scenarios where spatial representations of 3D objects are necessary, such as in 3D modeling \cite{Mine:2014:MVW} and medical imaging \cite{Zhang:2001:AIV}.



\subsection{Human Perceptual Psychology and Pyschophysics}
The sense of presence within VR draws on principles of Gibsonian psychology \cite{Gibson:1986:TEA} which ties human perception and movement to the overall comprehension of the environment. When the scene changes as result of a change in our head position, we perceive that as movement through an environment. 
Given the focus on the stimuli which simulate the physical measurements and quantities in data visceralization, we draw upon high level concepts from the extensive field of psychophysics \cite{Green:1966:SDT}, most notably the notion of human perceptual limits and how this may impact the ranges of stimuli used. We use this notion to systematically scope our design probes in the next section.

\DeclareRobustCommand{\EXOne}{%
  \begingroup\normalfont
  E1~\includegraphics[height=1.3\fontcharht\font`\B]{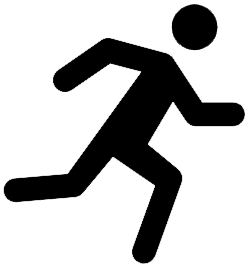}%
  \endgroup
}
\DeclareRobustCommand{\EXTwo}{%
  \begingroup\normalfont
  E2~\includegraphics[height=1.3\fontcharht\font`\B]{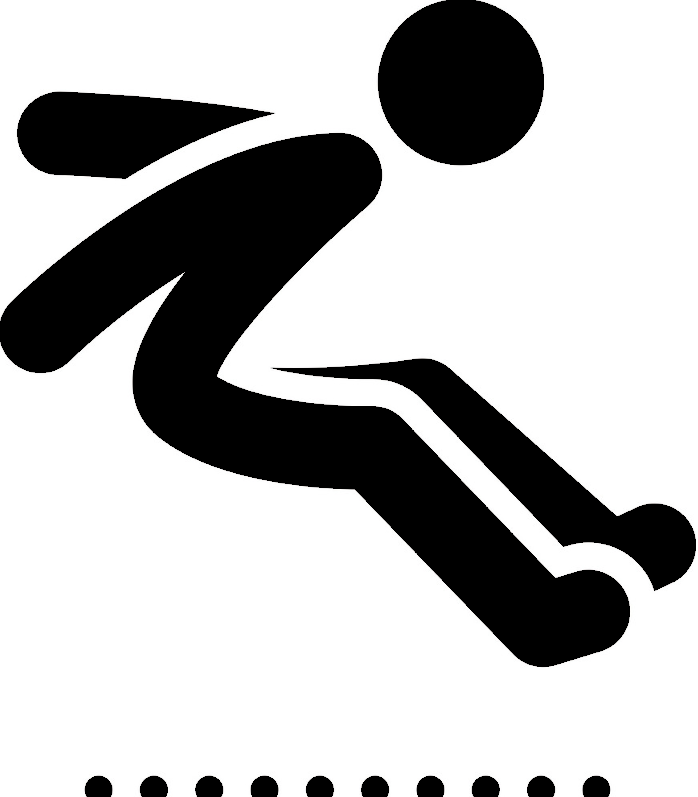}%
  \endgroup
}
\DeclareRobustCommand{\EXThree}{%
  \begingroup\normalfont
  E3~\includegraphics[height=1.2\fontcharht\font`\B]{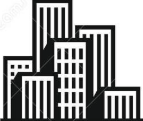}%
  \endgroup
}
\DeclareRobustCommand{\EXFour}{%
  \begingroup\normalfont
  E4~\includegraphics[height=1.15\fontcharht\font`\B]{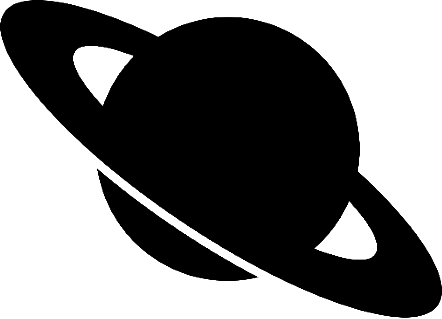}%
  \endgroup
}
\DeclareRobustCommand{\EXFive}{%
  \begingroup\normalfont
  E5~\includegraphics[height=1.15\fontcharht\font`\B]{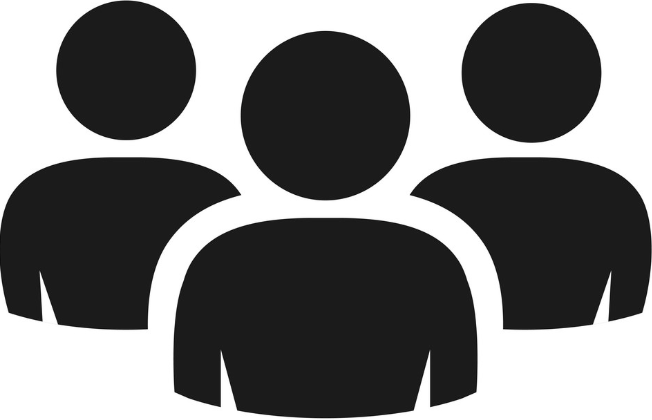}%
  \endgroup
}
\DeclareRobustCommand{\EXSix}{%
  \begingroup\normalfont
  E6~\includegraphics[height=1.15\fontcharht\font`\B]{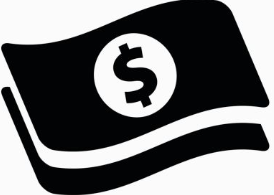}%
  \endgroup
}

\section{Design Probes into Data Visceralization} 
\label{sec:examples}
To explore and better define the concept of data visceralization, we developed a set of VR prototypes using the Unity3D game engine. \replaced{We critically reflect on these prototypes, with each prototype requiring 1 to 2 weeks to create, test, and critique.}{These prototypes were part of an iterative process \added{}, whereby the authors critically reflect on them.} 
These design probes were conducted using an ASUS Windows Mixed Reality Headset (HC102) in an open space free of obstructions. 
Each prototype was adapted either from existing stories published in online journals and news articles, or \replaced{from}{frp,} popular data-driven graphics/visualizations. We strove to test a range of different scenarios to explore data visceralizations as much as possible. In this section, we describe six of these design probes, which we refer to as examples for simplicity and abbreviate as \EXOne{}, \EXTwo{}, \EXThree{}, \EXFour{}, \EXFive{}, and \EXSix{}.
The first half focuses on scenarios with common types and scales of physical phenomena, not requiring any scaling or transformation to be readily perceived in VR. The second half investigates scenarios that required some form of transformation into VR, such as scaling down extreme values and representing abstract measures.
While we describe, show pictures, and include video of these examples, one needs to \emph{experience} them in a VR setup to assess the data visceralization experience. 
Therefore, we made these available in the supplementary material \replaced{and on a public GitHub repository}{(we will upload them on a publicly accessible website upon paper acceptance)}\footnote{https://github.com/benjaminchlee/Data-Visceralization-Prototypes}.

\begin{figure}
    \centering
    \includegraphics[width=\linewidth]{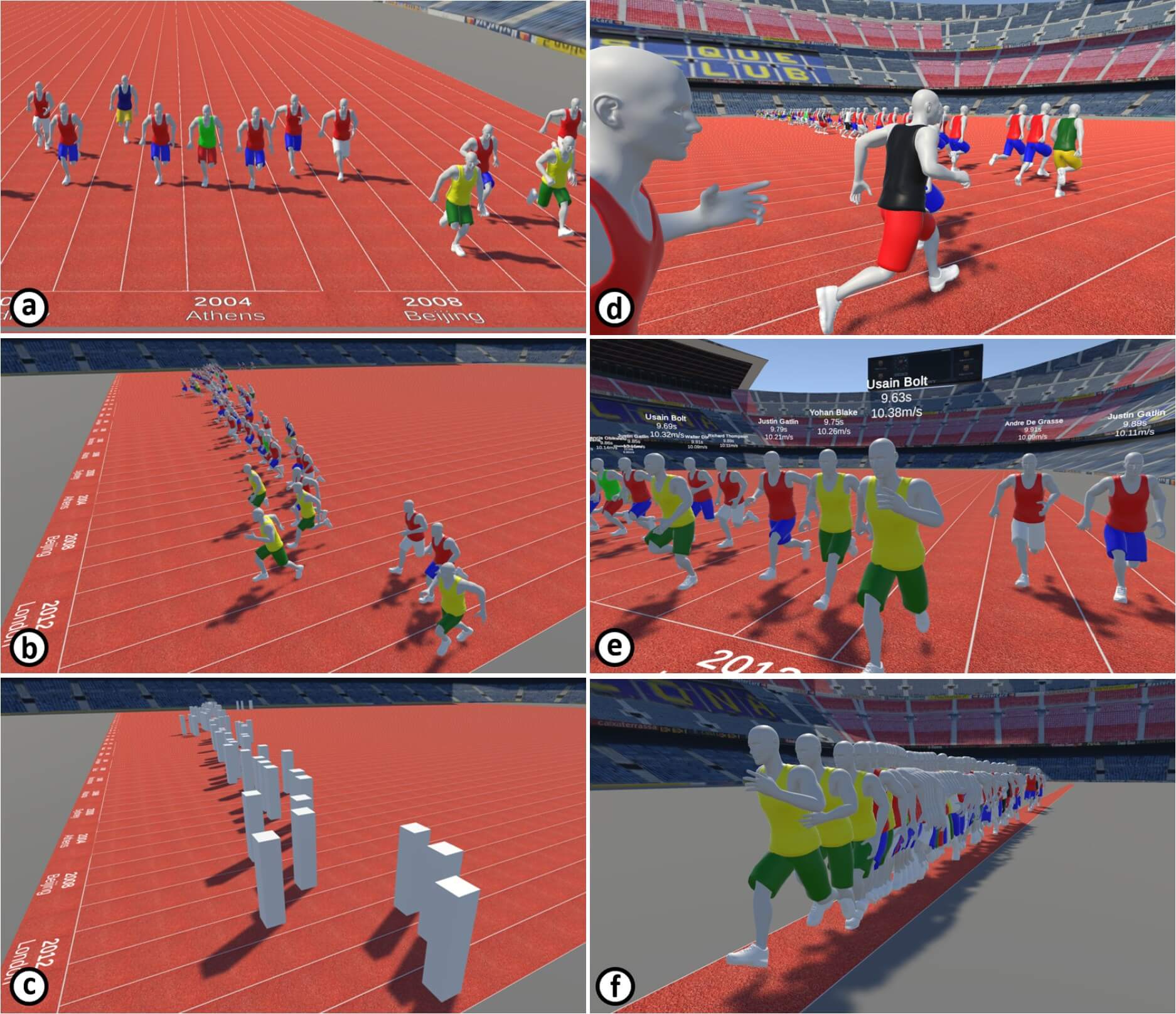}
    \caption{Illustrative images of \EXOne{}. (a) `Photo finish' of Olympic sprinters \replaced{(adapted from~\cite{NewYorkTimes:2012:Mens100m})}{taken from original data story's video \cite{NewYorkTimes:2012:Mens100m}}. (b) Overview of sprinters running down the track at real scale. (c) Same perspective but using human-sized cubes instead of humanoid models. (d) Experiencing the race from the perspective of the slowest sprinter. (e) Sprinters superimposed on top of one another. (f) Floating labels above each sprinter providing quantitative measures and values.}
    \label{fig:sprinters}
\end{figure}

\subsection*{\textbf{\replaced{E1}{EX1}}~\includegraphics[height=1.3\fontcharht\font`\B]{icons/runner.png} -- Speed: Olympic Men's 100~m}
Usain Bolt holds the world record in the Olympics Men's 100~m with his performance at the 2012 London Olympic Games at a time of 9.63~seconds. \textit{One Race, Every Medalist Ever} by the New York Times \cite{NewYorkTimes:2012:Mens100m} puts Bolt's result into perspective by comparing it with those of all other Olympic medalists since 1896. While this is commonly shown on a tabular layout (Fig.~\ref{fig:teaser}a), the story shows the relative distance away from the finish line each sprinter would be when Bolt finishes, highlighting the wide margins between him and the competition. This is shown in a video rendered with 3D computer generated graphics (Fig.~\ref{fig:sprinters}a) and an accompanying scatterplot-like visualization. This relative distance helps to quantitatively compare how much faster Bolt is than the rest. However, neither the visualization nor video give an accurate notion of just how fast Olympic sprinters can run. While watching it live in person may provide this, most people will not have the opportunity to do so, let alone stand on the track during the event, and it is impossible to resurrect runners from the last 100 years.

We based our prototype on the original story's video in a virtual environment at real-world scale (Fig.~\ref{fig:sprinters}b). The user can play, pause, and restart the race, causing each sprinter to run down the track at their average speed. We initially used simple human-sized cubes for the sprinters (Fig.~\ref{fig:sprinters}c), later replacing them with anatomical 3D models. We detail this notion later in Sec.~\ref{ssc:realism}. Thanks to the flexibility of VR, the race can be experienced from almost any perspective: standing at any position on the track and watching them run past, floating above the track, or even moving with the fastest/slowest sprinter (Fig.~\ref{fig:sprinters}d), which demonstrates the relative speed between sprinters and provides a glimpse of what it might be like from their perspective.
One clear issue early on was that by copying the original story's environment and its open, endless void, it was challenging to properly assess the speed of each sprinter. While it was still possible to make object-relative judgments between sprinters~\cite{Jerald:2015:VBH}, the lack of background meant there was no clear frame of reference for the virtual environment itself. We decided to add a stadium model around the track to resolve this issue, which also aided in immersion and contextual information to the experience. To further improve awareness, we also looked at adding optional annotations of exact times and speeds above each sprinter's head (Fig.~\ref{fig:sprinters}e), as well as experimenting with superimposing all sprinters on top of one another (Fig.~\ref{fig:sprinters}f). The latter was motivated by there being more than 80 lanes in the track, making it difficult to see everything at once. Note that we only re-scaled the width of the entire track for this, keeping the length of the track the same.


\subsection*{\textbf{\replaced{E2}{EX2}}~\includegraphics[height=1.3\fontcharht\font`\B]{icons/jump.png} -- Distance: Olympic Men's Long Jump}

\begin{figure}[th]
    \centering
    \includegraphics[width=\linewidth]{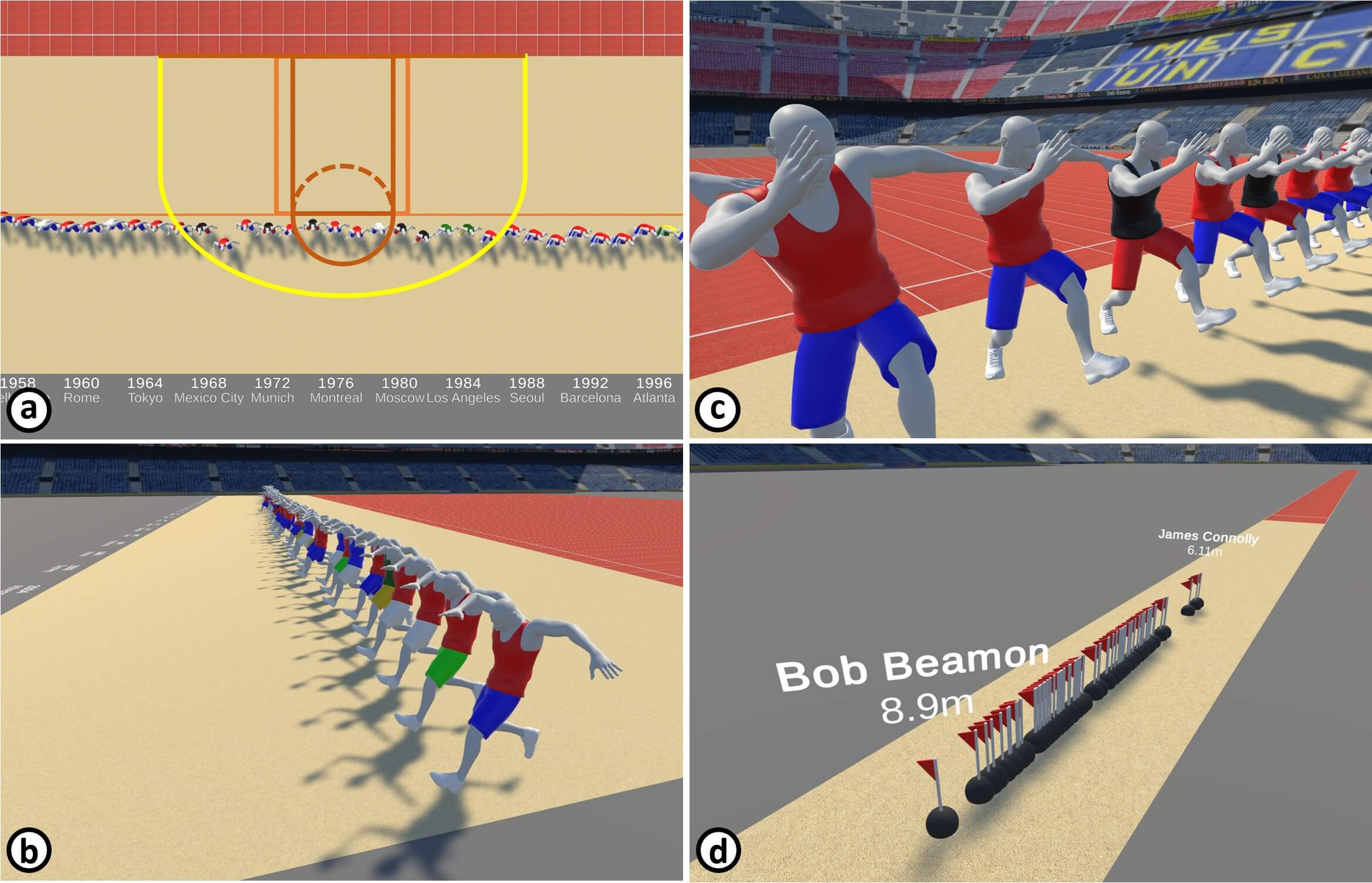}
    \caption{Illustrative images of \EXTwo{}. (a) Distances of long jumps compared to a 3-point line \replaced{(adapted from~\cite{NewYorkTimes:2012:MensLongJump})}{in the original data story's video \cite{NewYorkTimes:2012:MensLongJump}}. (b) Overview of long-jumpers at the final `freeze-frame' when landing. (c) Experiencing the jump as it happens from a close angle. (d) Long-jumpers represented as \added{superimposed} flags on the ground\deleted{, superimposed together} with labels on the furthest and closest jumps.}
    \label{fig:longjump}
\end{figure}

Unlike the progressively record breaking times of the Olympic Men's 100~m into the new millennium, 1968 saw Bob Beamon get the world record for the farthest long jump at 8.9~m (29~ft 2.5~in), and still holds the \textit{Olympic} record to this day---losing the \textit{world} record to Mike Powell in 1991 at 8.95~m (29~ft 4.25~in). Told in a similar fashion to the Men’s 100~m Sprint story with video and visualization~\cite{NewYorkTimes:2012:Mens100m}, \textit{Bob Beamon's Long Olympic Shadow}~\cite{NewYorkTimes:2012:MensLongJump} highlights not only Beamon's performance relative to his peers, but also the sheer distance that these athletes can jump to begin with: comparable to the distance of a basketball 3-point line (Fig.~\ref{fig:longjump}a). This comparison serves to put these distances into perspective, but what if one hasn't played basketball or even set foot on a basketball court? Of course, it is possible to use a more familiar anchor instead~\cite{Hullman:2018:ICM}, 
but visualizing these distances at real-life scales would allow users to experience them for themselves.

Our VR prototype shares much in common with \EXOne{}, the key difference being the use of distance rather than speed. While the user can still control the playback of the event, the focus is on the final `freeze-frame' of each long jumper at the end of their jump (Fig.~\ref{fig:longjump}b). The user can change their viewpoint to any position to get a visceral sense of the sheer distance of the long-jumpers (Fig.~\ref{fig:longjump}c), which is not possible by just watching the video. 
As is the case with \EXOne{}, we also superimpose all of the long-jumpers on top of one another for easier comparison. Given the precision and fine differences of the results in long jump however, the long-jumpers can instead be viewed as planted flags to more closely judge distances between each other (Fig.~\ref{fig:longjump}d).


\subsection*{\replaced{E3}{EX3}~\includegraphics[height=1.3\fontcharht\font`\B]{icons/height.png} -- Height: Comparison of Skyscrapers}

When comparing physical measurements of related objects or creatures, one common visualization to use is a comparison diagram. This juxtaposes each subject as 2D images or silhouettes, with the y-axis usually showing height. One popular use of this is the comparison of skyscrapers, such as the one shown in Fig.~\ref{fig:teaser}c. These are great for understanding relative sizes between objects (e.g., that Mount Elbrus, Kilimanjaro, and Denali are roughly equivalent in elevation), but not the absolute size of each subject (i.e., what it feels like to be at the base of a mountain roughly 6~km or 3.7~mi tall).

\begin{figure}
    \centering
    \includegraphics[width=\linewidth]{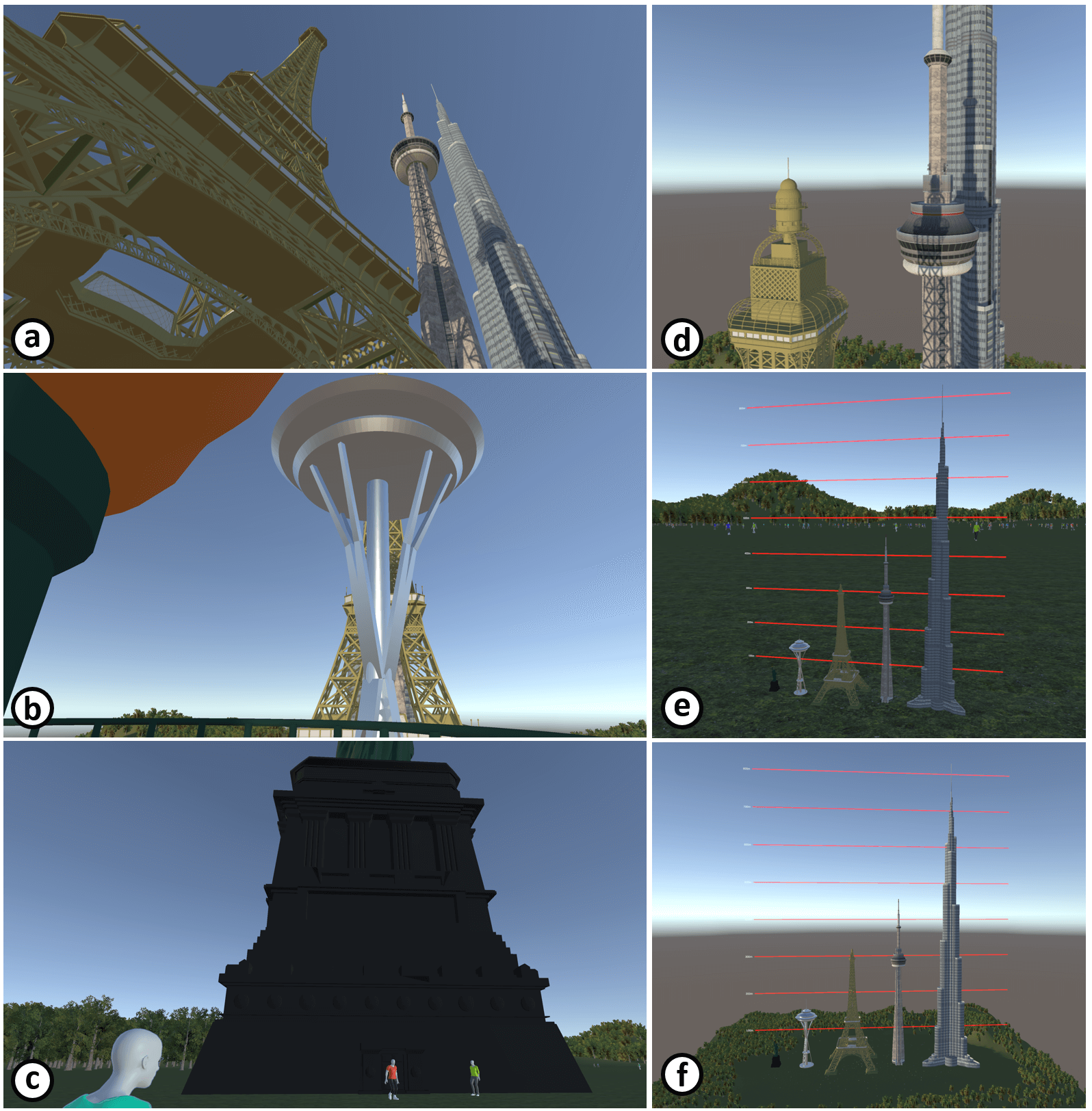}
    \caption{Illustrative images of \EXThree{}. (a) Looking at the Eiffel Tower, CN Tower, and Burj Khalifa at real-life scale from below. (b) Looking at the Space Needle from the top of the Statue of Liberty. (c) Crowds randomly walk around on the ground floor. (d) Shadows cast from one building to another from a light source directed parallel to the ground. (e) Miniaturized 3D version of comparison diagram with y-axis scale. (f) Same setup but from a distant view.}
    \label{fig:skyscrapers}
\end{figure}

We decided to create a VR version of the skyscraper diagram seen in Fig.~\ref{fig:teaser}c. We chose skyscrapers because their size can be overwhelming to see and experience, but is still familiar to people living in cities with tall skylines. Our prototype uses 3D models of famous skyscrapers and landmarks (Statue of Liberty, Space Needle, Eiffel Tower, CN Tower, and Burj Khalifa) positioned side-by-side in a similar fashion to the original visualization. These models are scaled as accurately as possible to their real-world counterparts. As with the prior examples, it is possible to move around the environment in ways that are either not available to most people, such as viewing from below (Fig.~\ref{fig:skyscrapers}a) or from the top of each skyscraper (Fig.~\ref{fig:skyscrapers}b), or in ways that are physically impossible, such as flying around in a `Superman'-like fashion.
We also experimented with adding visual cues to aid in scale perception, such as life-sized people randomly walking at ground-level (Fig.~\ref{fig:skyscrapers}c) to simulate the feeling of `seeing people as ants,' and casting shadows from skyscrapers to the ground (Fig.~\ref{fig:skyscrapers}d) as an artificial `ruler.' In addition, we explored two alternative views of the scene. The first was to miniaturize the skyscrapers such that they were approximately 2~m (6~ft 7~in) in size and allowing the user to pick up and re-position them (Fig.~\ref{fig:skyscrapers}e). The second was to retain their real-world scale, but position the user far away enough so that the skyscrapers still occupied a similar space in the user's field of view (Fig.~\ref{fig:skyscrapers}f). We elaborate on this difference in Sec.~\ref{ssc:mini-vs-distant}.


\subsection*{\textbf{\replaced{E4}{EX4}}~\includegraphics[height=1.3\fontcharht\font`\B]{icons/system1.png} -- Scale: Solar System}
\begin{figure}[bh]
    \centering
    \includegraphics[width=\linewidth]{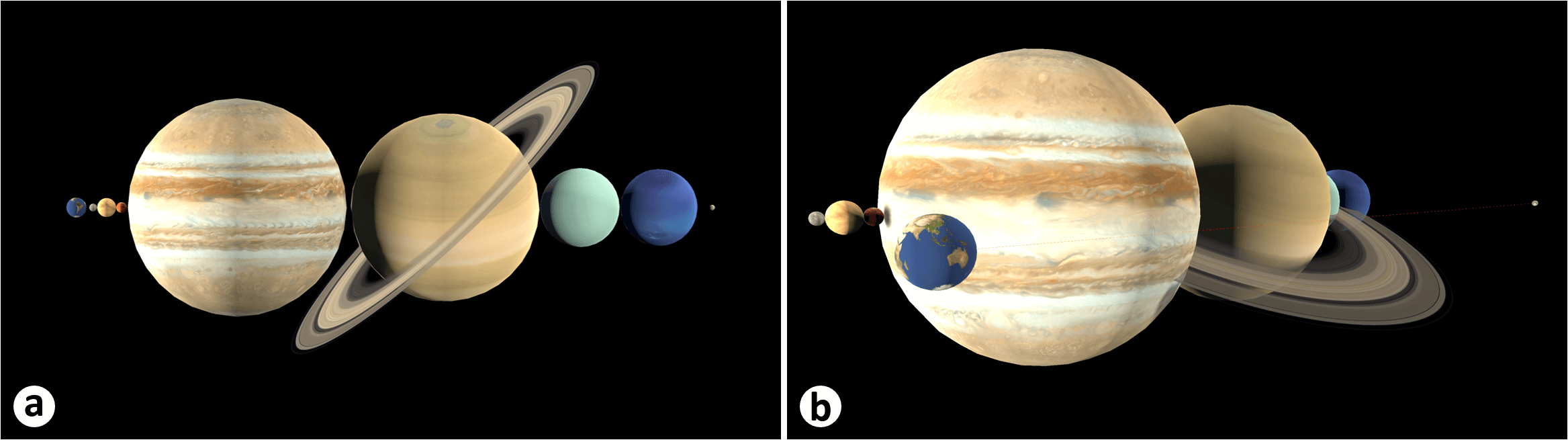}
    \caption{Illustrative images of \EXFour{}. (a) \replaced{Reference scene that our prototype is based on (adapted from~\cite{PerplexingPotato:2014:Planets})}{Original image used to base the prototype off of \cite{PerplexingPotato:2014:Planets}}. (b) Prototype version in VR at a 1:40,000,000 scale.}
    \label{fig:solarsystem}
    \vspace{-3mm}
\end{figure}

Many graphics, visualizations, and videos exist to educate people on the enormous scale of our solar system. One such example~\cite{PerplexingPotato:2014:Planets} (Fig.~\ref{fig:solarsystem}a) focuses on the surprisingly large distance between the Earth and the Moon. It does so by illustrating how all of the other planets can fit between the two when at their average distance apart. This presented a challenge which we wanted to investigate: the effects of re-scaling and transformation on viscerality in VR.

In our prototype (Fig.~\ref{fig:solarsystem}b), we chose a ratio of 1:40,000,000 (i.e., 1~m in VR = 40,000~km in space), resulting in a real-world distance of approximately 10~m between the Earth and Moon. This struck a balance between being small enough to be able to see all planets reasonably well, but large enough to still be at a reasonable size (Earth at roughly 30~cm (12~in) in diameter). As the intent is to see the other planets comfortably fitting between the Earth and Moon, the user can grab and slide these planets as a group along a single axis to align them all together. Note that for \EXFour{}, \EXFive{} and \EXSix{}, we detail these challenges in transformation in Sec.~\ref{ssc:data-transformation}.

\begin{figure}
    \centering
    \includegraphics[width=\linewidth]{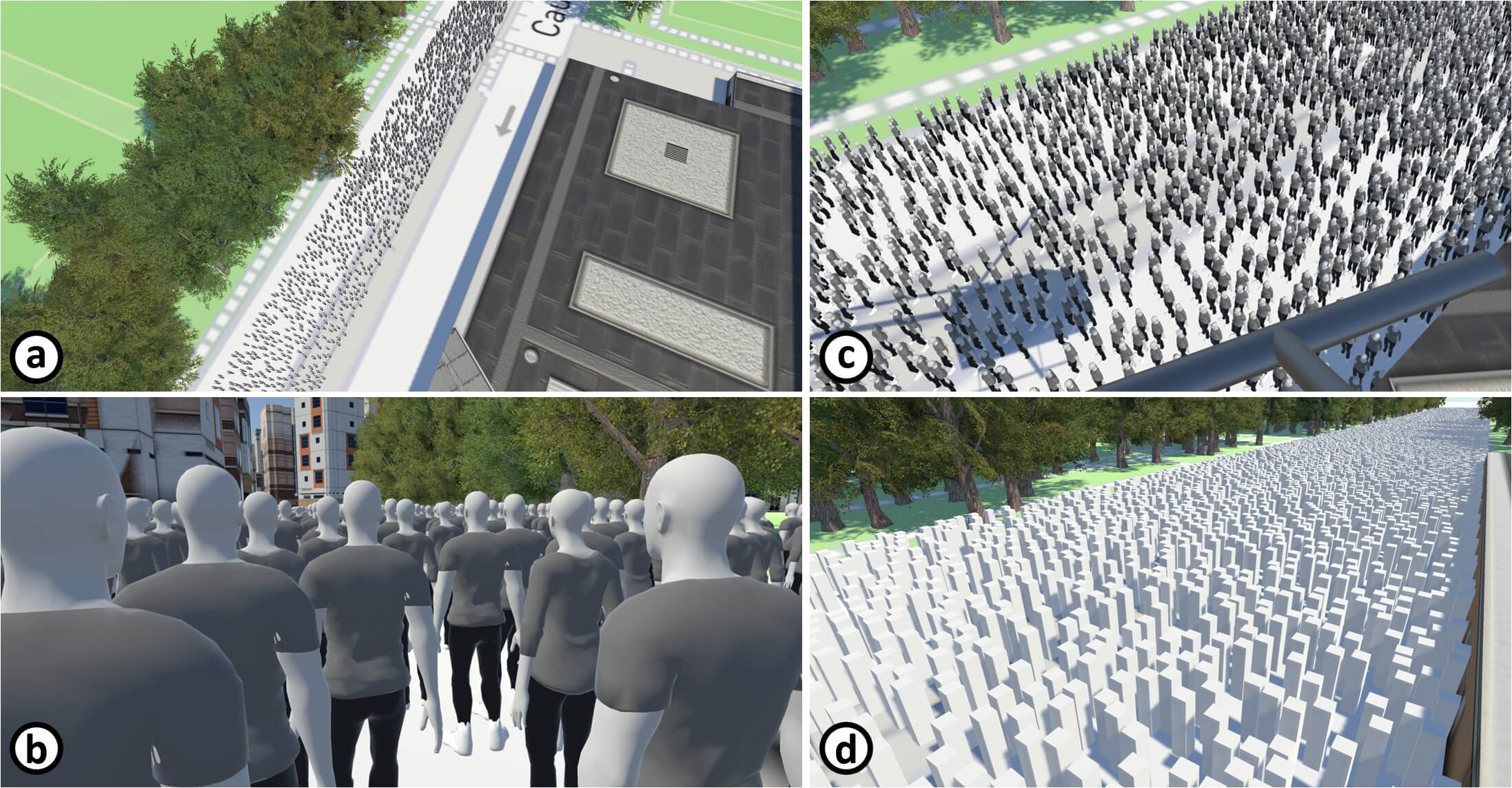}
    \caption{Illustrative images of \EXFive{}. (a) \replaced{Birds-eye view of the protest (adapted from \cite{NewYorkTimes:2019:HongKongProtest})}{Screen capture of a portion of the original story \cite{NewYorkTimes:2019:HongKongProtest}}. (b) Perspective from inside the crowd. (c) Perspective from above while flying in a helicopter-like fashion. (d) Crowd represented as human-sized cubes.}
    \label{fig:protests}
\end{figure}

\subsection*{\textbf{\replaced{E5}{EX5}}~~\includegraphics[height=1.3\fontcharht\font`\B]{icons/crowed.png} -- Discrete Quantities (of Humans): Hong Kong Protests}
Protest marches on Hong Kong's streets beginning in 2019 involved over 2 million people. To help visualize the sheer scale of this event, the New York Times stitched together several birds-eye view photographs from June 16, 2019 to form a vertical composite image in a scrolling format~\cite{NewYorkTimes:2019:HongKongProtest} (Fig.~\ref{fig:protests}a). It combines both the small size of each individual person with the seemingly never-ending photographs to accomplish this. This begs the question, however, what is it actually like to be there in person? Is it a noticeably different experience being able to walk throughout a large crowd of people as compared to looking at photos alone, and if so, can we perceive the extreme quantities on display?

Given the technical limitations of rendering upwards to a million animated 3D humanoid models in VR, we recreate only a small section of the protest using 10,000 models. The crowd and surrounding environment are at a real-world scale, and it is possible to move through the crowd to experience being surrounded by many people (Fig.~\ref{fig:protests}b). The user can also choose to fly above the crowd in a similar way to a helicopter, without any physical limitations (Fig.~\ref{fig:protests}c). However, it was clearly apparent that the idle nature of the protesters detracted a lot from the experience, to the point where using static human-sized cubes in place of the models worked just as well to convey quantities (Fig.~\ref{fig:protests}d). We suspect that much more varied models, animated movements, and lighting would go a long way to provide a more visceral experience of being at the protest.


\subsection*{\textbf{\replaced{E6}{EX6}}~\includegraphics[height=1.3\fontcharht\font`\B]{icons/bill.png} -- Abstract Measures: US Debt Visualized}
\begin{figure}
    \centering
    \includegraphics[width=\linewidth]{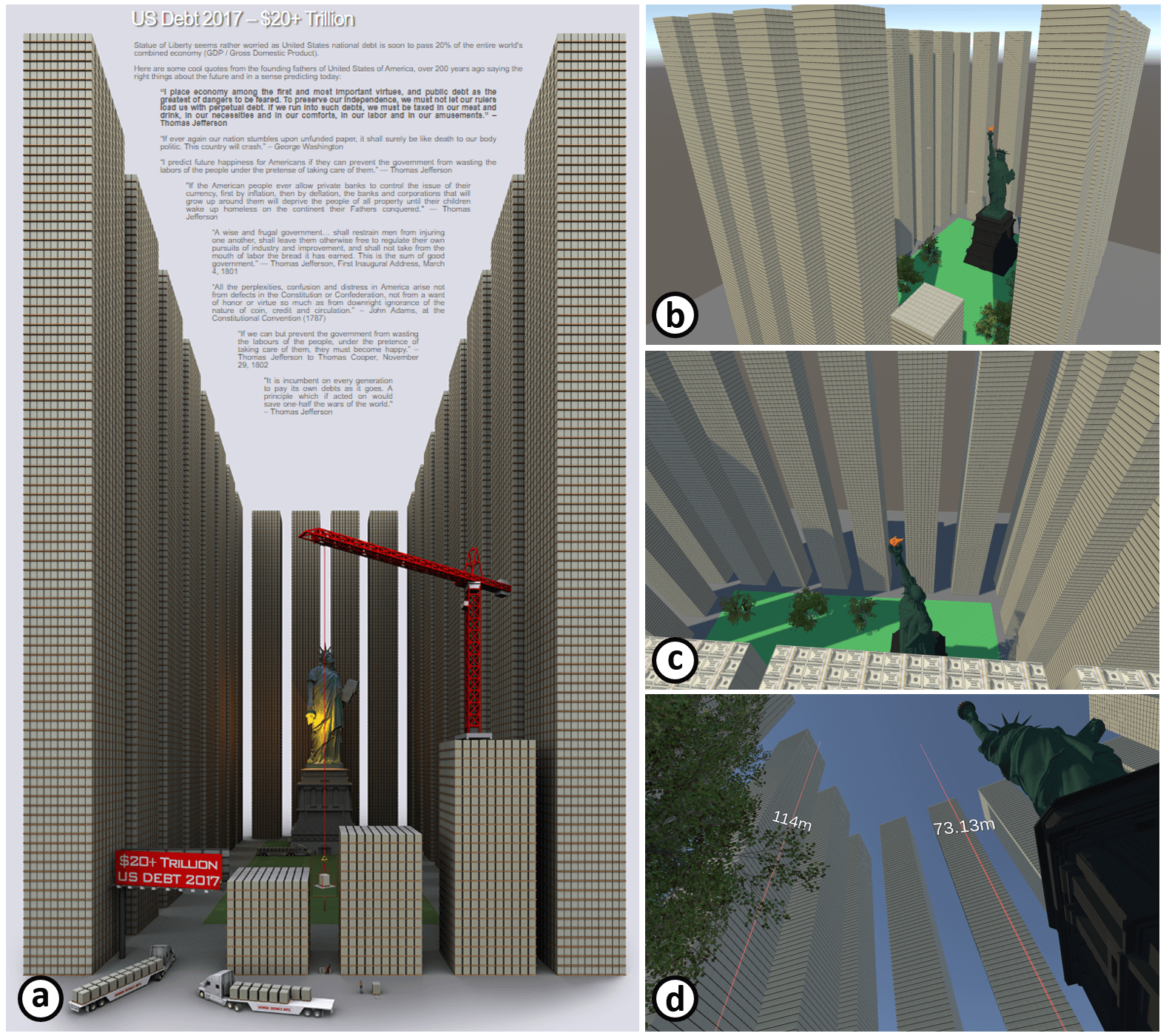}
    \caption{Illustrative images of \EXSix{}. (a) \replaced{Reference scene that our prototype is based off (\textcopyright{} Oto Godfrey, Demonocracy.info, used with permission)}{Screen capture of the final section of the original story \cite{Demonocracy:2017:USDebt}}. (b) Overview of the scene with the Statue of Liberty in the center. (c) Looking down from the top of a stack. (d) Looking up from ground level and at labels of stack and Statue height.}
    \label{fig:usdebt}
    \vspace{-3mm}
\end{figure}

In \textit{US Debt Visualized in \$100 Bills} \cite{Demonocracy:2017:USDebt}, an incomprehensibly large amount of money (\$20+ trillion USD in 2017) is visualized using concrete scales. Starting from single \$100 bills and moving to pallets of bills worth \$100 million each, the piece culminates in comparing stacks of these pallets with other large objects, such as the Statue of Liberty as seen in Fig.~\ref{fig:usdebt}a, putting into perspective the amount of debt. However, the true scale of each stack is difficult to properly grasp from the image, let alone without having visited the Statue of Liberty or stood under a construction crane. A similar concept was already investigated in \EXThree{}, but as money is inherently conceptual, would this transformation from abstract measure to physical measure (i.e., size of \$100 bills) influence viscerality?

Our VR prototype closely replicates the original piece, with stacks of pallets of bills surrounding the Statue of Liberty (Fig.~\ref{fig:usdebt}b) with an updated total of \$22+ trillion USD. Each stack is comprised \deleted{out }of $10\times10\times100=10,000$ life-sized pallets of \$100 bills, piling up to approximately 114~m in height. Given both this and \EXThree{} primarily convey height, they offer similar points of view, such as from the top of a stack (Fig.~\ref{fig:usdebt}c) or from the bottom of a stack (Fig.~\ref{fig:usdebt}d). To try and quantify some aspects of the experience however, we add annotations of certain heights of the objects in the scene (Fig.~\ref{fig:usdebt}d).

\section{Critical Reflection on Design Probes} \label{sec:reflection}
In this section, we go into detail of our observations and critiques of the design probes described in the previous section. 


\subsection{Perception: Perceptual `Sweet-spots'} 
\label{ssc:sweet-spots}
It is clear from \EXOne{} and \EXTwo{} that VR experiences of human-scale data fall into a perceptual sweet-spot for data visceralization. Given that these directly relate to human performance in sport, measures such as running speed and jump length can directly be experienced with VR. \EXThree{} began to highlight some of these limits however, as all skyscrapers simply appear to be tall beyond a certain height, \replaced{requiring special consideration}{meaning that additional steps needed to be taken in order} to mitigate this (Sec.~\ref{ssc:virtuality}).
However, a given measure being within this sweet-spot did not automatically make it easy to understand \replaced{every detail of the data}{without any faults}. For example, while \EXTwo{} still conveyed the sense of \deleted{long jump }distance, the\deleted{se} values were very similar to each other, making it difficult to make comparisons between athletes.



\subsection{Virtuality: Manipulating the Scene to Facilitate Understanding} \label{ssc:virtuality}
Because we designed and implemented our prototypes using VR, the virtual world allowed us to manipulate both the virtual objects and the user's viewpoint in ways not possible in real-life in order better facilitate understanding of the chosen phenomena.
One example of manipulating objects was the positioning of athletes in \EXOne{} and \EXTwo{}. Perspective foreshortening can distort the relative perception of objects in the environment, meaning that comparing athletes on either end of the $\approx$80~m track was difficult. In the original videos \cite{NewYorkTimes:2012:Mens100m, NewYorkTimes:2012:MensLongJump} the use of fixed view points, lack of stereopsis, and orthogonal perspectives avoided this issue. Conversely, the use of VR allowed us to manipulate and play with the position of the athletes, superimposing them on top of one another (Fig.~\ref{fig:sprinters}f) to achieve a similar orthogonal view, therefore making this comparison easier.
Likewise, we can manipulate the position of the user and their viewpoint to make it easier to view and understand the phenomena, as can be seen in \EXThree{}. Certain viewpoints made it difficult to accurately judge and/or make comparisons between skyscrapers, such as when standing near the base of a skyscraper and looking directly straight up, or when the skyscraper was self-occluding (e.g., the Space Needle's observation deck). However, we overcome this by allowing the user to fly and teleport around the scene, going to the top of any skyscraper they wished. Doing so reduces the distance and angle to the other skyscrapers, making it easier to see.
A surprising side effect this however, was that it provided more nuanced insights in the data, such as in \EXOne{} where moving along with the fastest/slowest sprinter grants both an experience of the speed that they were moving at, but also to compare and contrast their speed directly with all the others.
More broadly, this concept of manipulation to facilitate understanding is prevalent throughout our design probes, such as juxtaposing skyscrapers from different continents in \EXThree{} and moving all planets closer together in \EXFour{}. However, if the goal is strictly to viscerally understand the data in a manner as close to reality as possible, this may negatively impact that understanding as these manipulations may be deemed unrealistic and off-putting to users.

\subsection{Realism: The Role of Photorealism and Abstraction} \label{ssc:realism}
In all of our examples, we needed to add relevant contextual backgrounds and visual aids, such as the stadium in \EXOne{} and \EXTwo{}\deleted{, as well as the crowds in \EXThree{}}. The lack of contextual cues in the environment detracted from the sense of presence, and in certain cases made the data more challenging to understand. In \EXOne{}, for example, moving along with a sprinter without any background environment present removed all sense of absolute motion, whereas in \EXThree{} and \EXSix{}, not having a floor made it difficult to judge overall height. 
We also explicitly manipulated the level of realism of the physical phenomena in several of the examples, such as using cubes instead of humans in \EXOne{} and \EXFive{}, as well as small flags in \EXTwo{}. Surprisingly, this had little impact on the visceral feeling of the data---a human-sized cube moving as fast as an Olympic sprinter still conveys the speed in a similar manner. These observations aside, it is still unclear what role realism plays. A basic level of detail of background environments may be necessary, but it does not need to be high-fidelity. More photorealistic rendering and simulation may be more engaging and enjoyable, but conveying visceral senses of motion and distance were achievable without highly sophisticated representations.


\subsection{Annotation: The Use of Annotations and Distractions} \label{ssc:annotation}
While our senses are very good at judging relative sizes and speeds, \replaced{we}{our direct senses} are less well suited to determining absolute values---hence the need \replaced{to display exact}{for some sort of display of the} measurements. In several examples, we experimented with augmenting the direct experience with annotations to show sizes or distances. We made sure these could be turned on and off, since we were concerned that such augmentation might impact the realism and thus the direct perception of the measures. However, as in more primitive representations, we saw little detraction from the use of annotation. This was moderately surprising since, as Bertin specifies \cite{Bertin:1983:SG:1095597}, reading text can capture user attention and thus reduce the perception of other stimuli, negatively impacting the visceral experience with the data.


\subsection{Knowledge Transfer: Applying Knowledge and Experience from Visceralization into Real-Life Contexts} 
\label{ssc:knowledge-transfer}
Data visualisations share information and insights in data. In contrast, the physical nature of visceralizations may convey a different type of understanding that is more grounded in reality. For instance, after having seen the skyscrapers in \EXThree{}, one may then be able to `transfer' that knowledge to real-life skyscrapers in their day-to-day life, comparing this new experience with the prior VR one. Likewise, one may see the skyscrapers in VR and be reminded of previous real-life experiences. While similar in premise to VR for tourism \cite{Guttentag:2010:VRA}, the nature of visceralizations being more closely tied to data may aid users' perception and understanding of these physical phenomena in the wild.


\subsection{Data Transformation} \label{ssc:data-transformation}
Given that many phenomena we wish to understand fall outside of the realm of direct perceptibly, parts of \EXThree{}, \EXFour{}, \EXFive{}, and \EXSix{} helped illuminate various pitfalls that may occur when scaling large data into ranges more readily understandable by viewers, or when remapping abstract measures into perceivable concrete units.

\begin{figure}
    \centering
    \includegraphics[width=0.95\linewidth]{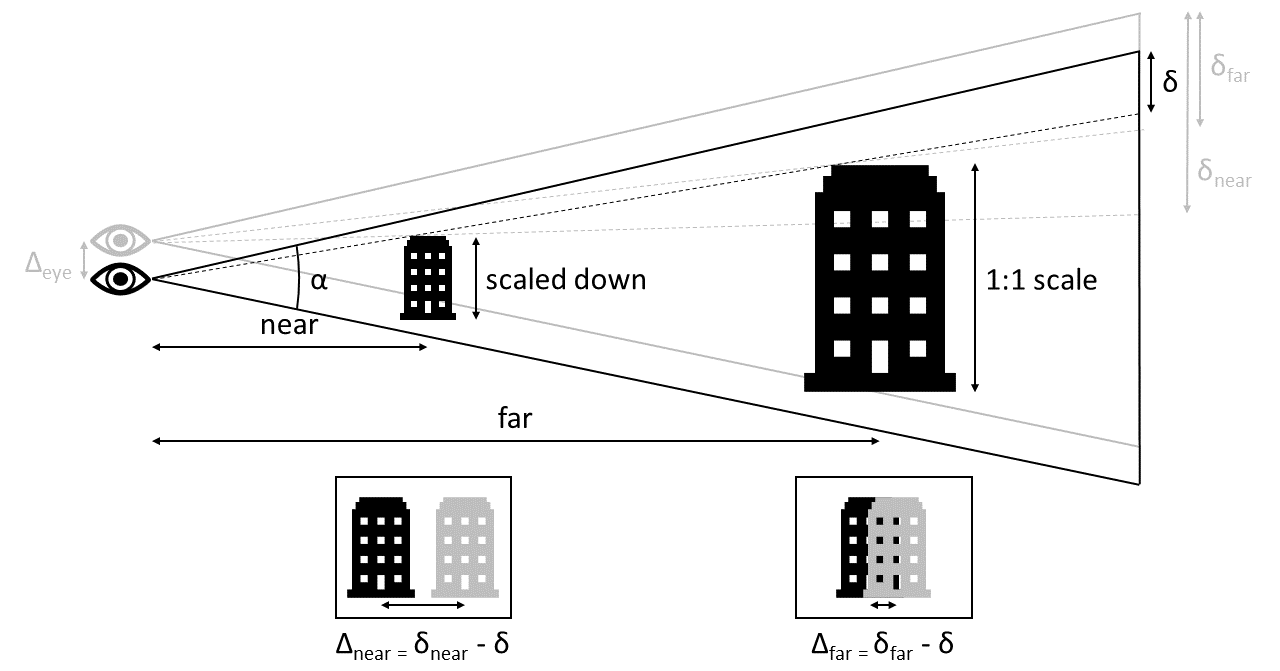}
    \caption{
    Illustration of the difference in perception when viewing an object at two different scales and distances, such that the object fills the same apparent size in the field of view. Consider a given point on the object which is projected onto a position of distance $\delta$ from the edge of the field of view of angle $\alpha$. When the eye is moved by a distance $\Delta$eye, this point in the scale-down object is shifted by a distance $\Delta$near=$\delta$near$-\delta$ and for the real-scale object $\Delta$far= $\delta$far$–$$\delta$. $\Delta$ far is much smaller than $\Delta$ near.
    }
    \label{fig:scaledvsdistant}
    \vspace{-3mm}
\end{figure}

\bpstart{Miniaturization: Maintaining viscerality during scaling} \label{ssc:mini-vs-distant}
We chose to use skyscrapers in \EXThree{} as we considered them on the cusp of being both perceivable but also somewhat too large to properly see. To this end, we experimented with having: (1) a scaling transformation applied to have a miniaturized view of the skyscrapers; and (2) a distant yet visibly equivalent view of the skyscrapers, retaining absolute scales. Despite appearing the same on a traditional monitor, they had very noticeable differences in their visceral nature when in VR. The first instance produced a significant perceptual mismatch, where motion parallax (caused by the motion of our head or body) caused significant perceived motion of the scaled skyscrapers, but little perceived notion in the second instance when far away (Fig.~\ref{fig:scaledvsdistant}). 
While the latter looked to be real skyscrapers from really far away, the former looked like miniature 3D printed models which broke the illusion of being real skyscrapers. Similarly, photographers have used perceptual mismatch techniques to make real photographs look like miniature models by adjusting angle and blur to simulate a limited focal length exposure in tilt-shift photography.

\bpstart{Extreme Scales: Experiencing phenomena significantly out of range of perception} 
\label{ssc:out-of-perception}
\EXFour{} explored a scenario where visualizing the data at a one-to-one scale would be impossible through the size of the planets. As a result of the scaling, much of the experience became similar to the miniaturized skyscrapers, where they clearly looked like scale models. Because of this, it was impossible to get any direct notion of the true size of any of the planets. However, as the relative sizes between the planets and the relative distance between the Earth and Moon were preserved, it was still possible to make comparisons between the planets and the Earth--Moon gap---arguably the intention of the original piece regardless.
\EXFive{} instead explored very large quantities of discrete objects, each one at a human perceptible scale (i.e. a large crowd). While the intention was to be `lost' in the `sea' of people, what we found was that the occlusion of nearby people made it challenging to see long distances, and in cases where this was possible, perspective foreshortening shrunk the heads of those further away. This resulted in a metaphorical `bubble of perception', where one could only see and get a feel for the number of people within the bubble. Conversely, density was quite simple to gauge as it only relied on estimation of the immediate surroundings. For anything involving the larger crowd as a whole (i.e. tens of thousands), changing viewpoints to have a vantage point above the rest is necessary, however at certain elevations it becomes very similar to watching a news broadcast.

\bpstart{Abstract Values: Perceiving abstract values with data visceralization} 
\label{ssc:abstract-concepts}
\EXSix{} explored the notion of requiring some transformation from abstract concept to concrete object, in this case mapping money of the US debt to \$100 bills. While the sheer scale of the stacks of money was present in a fashion similar to \EXThree{}, there was a level of cognitive effort required to mentally translate the visceral understanding into the abstract quantity, such that it was difficult to get any sense of direct, deep understanding of money. Since there was already a transformation from quantity of dollars into stacks of bills, it is unclear whether the VR representation gave any deeper understanding of said quantity more than the original 2D illustration would. In a sense, while the visceral experience of the concrete, physicalized representation was there, there was no visceral understanding of the original quantity itself.

\section{User Feedback} 
\label{sec:user-study}
To further expand our critical reflection and minimise bias, we invited external participants to try our prototypes and give feedback on the concept of data visceralization. Since it remains a novel and somewhat abstract concept, we chose not to formally measure and evaluate the effectiveness of data visceralization, instead focusing on participants' thoughts and opinions which will help us better define the concept.


\subsection{Process}
\begin{wrapfigure}[3]{r}{0.22\linewidth}
\hspace{-0.3cm}
\includegraphics[trim = 0 0 0 440, width=1.05\linewidth]{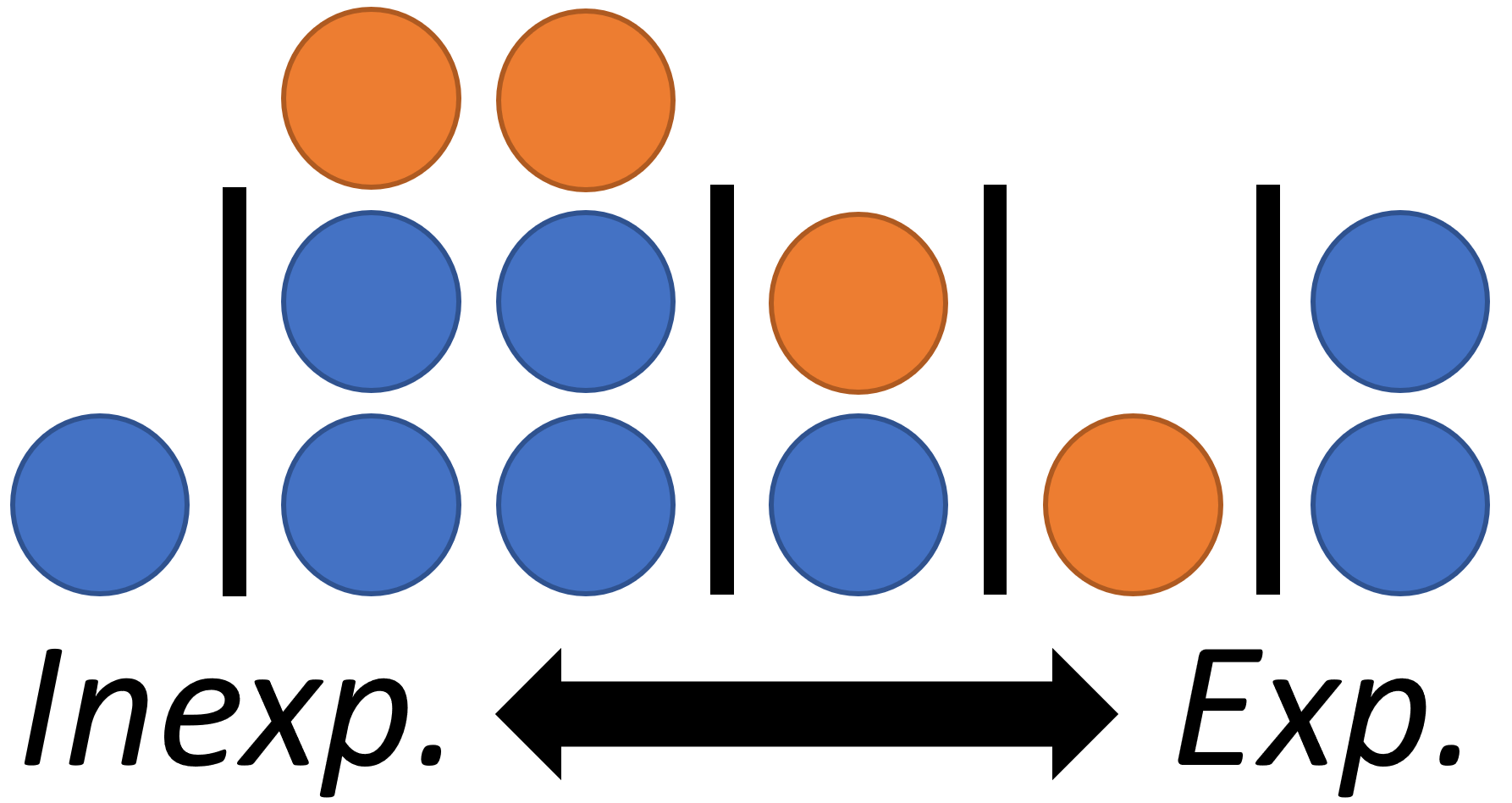}
\end{wrapfigure}
We recruited 12 participants (4 female \includegraphics[height=0.8em]{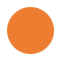}) who were all university students with a range of different backgrounds (majority computer science) and experience with VR, as indicated to the right. They were not given any monetary compensation for their time.
We conducted the sessions in a large open space free of obstructions using a HTC Vive Pro connected to a desktop PC with an Intel Core i7-7800X CPU (3.5~GHz, 6 cores), Nvidia GeForce GTX 1080 (8~GB) GPU, and 32~GB of RAM. We limited the scope to three of the six prototypes described in Sec.~\ref{sec:examples}: \EXOne{}, \EXThree{}, and \EXSix{}, conducted in that order \added{from simplest to most complex, however the lack of randomization may had influenced user preferences}. This was \added{done} due to time constraints, but still allowed us to \replaced{obtain}{attain} feedback from a representative sample of measures: speed, scale, and abstract quantities.
For each prototype, participants were given both the original source material and data visceralization to try (which we now refer to as desktop version and VR version respectively). For equivalency, we trimmed the desktop version to its relevant parts and mute any sound and voiceover.
Each was followed by a questionnaire (with Likert scales from 1 (strongly agree) to 7 (strongly disagree)) asking for their thoughts and opinions between the two versions. Note that the goal was not to measure which version was better, but to understand their strengths, weaknesses, and characteristics.
Sessions were concluded with a semi-structured interview to elicit detailed responses and opinions of data visceralization. This was loosely structured around the topics discussed in Sec.~\ref{sec:reflection}, but was conducted in such a manner to allow for broader discussion driven by the participants.

The questionnaire responses and de-identified transcripts are available in supplementary material. We combined interview notes and qualitative analysis on the transcripts to identify common and interesting themes. We include only the most relevant quotes below, labeled by participant. Please refer to the transcripts for additional context.

\subsection{Findings} \label{ssc:user-study-findings}
We first report on high-level results and metrics, followed by insights categorized and contrasted against relevant topics \added{that were} described in Sec.~\ref{sec:reflection}. We then discuss themes raised by our participants which were not part of our reflection later in Sec.~\ref{sec:discussion}.

\subsubsection{General Results} \label{ssc:general-results}
\begin{wrapfigure}[10]{r}{0.285\linewidth}
\hspace{-0.325cm}
\includegraphics[trim = 20 30 0 115, width=1.1\linewidth]{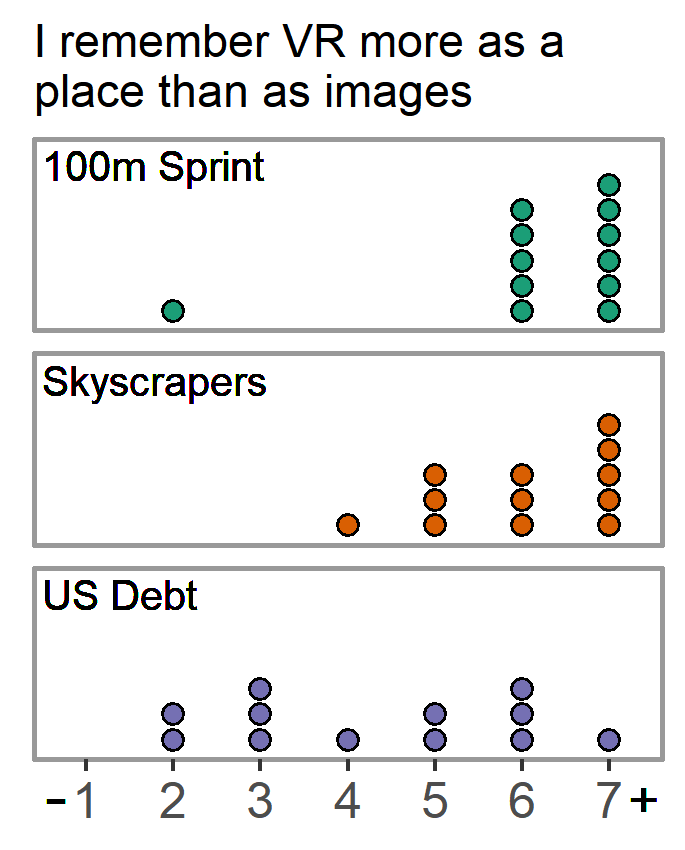}
\end{wrapfigure}
Each session lasted for an average 82.4 minutes (\textit{Mdn} = 85.5, \textit{SD} = 13). Overall, all participants preferred VR for \EXOne{} and \EXThree{}, but were mixed between desktop (2 preferred), VR (6), and no preference (4) for \EXSix{}. 
Similarly, they reported to be more immersed in the first two prototypes, particularly due to the `unrealistic' stacks of \$100 bills being off-putting. Participants generally liked the greater freedom and immersiveness of VR, but criticized the poor communication of numerical values, specifically for \EXSix{} (discussed later). Participants spent more time on average in VR than the desktop version: 3.4 times the duration for \EXOne{}, 13.2 times for \EXThree{}, and 1.8 times for \EXSix{}.
\added{Cybersickness in VR was reported by a minority of participants only when they used flying locomotion techniques, whereby they were advised to use other techniques instead (i.e., teleportation).}
Most participants chose to watch the desktop versions of \EXOne{} and \EXSix{} once (presented as videos) with a few watching it twice. All participants were familiar with \EXThree{} (presented as an image).

\subsubsection{Comparison to Critical Reflection}
\bpstart{Virtuality: Freedom of changing viewpoint provides unique insights but requires more time to do so}
In Sec.~\ref{ssc:virtuality} we considered how the manipulation of the virtual environment can facilitate understanding but negatively impact viscerality.
While no participant specifically mentioned that this aided their understanding,
three participants remarked that having these types of environments is a core part of VR as it \textit{``Allows you to do something that you otherwise would never see.''} [P10].
In contrast, two other participants stated that being in these unrealistic scenarios made them feel `out of place', such as when flying around the environment.
Neither participant said that this negatively impacted their understanding of the phenomena, however one of them clarified that things still need to be \textit{``Presented in a way that hypothetically could exist.''} [P11], such as stacks of \$100 bills technically being possible to create if one had enough resources.
In terms of the user's viewpoint, many of the benefits we described were also identified by participants,
particularly the ability to view the data from any angle they wished.
This meant \textit{``Every angle I was at, I could take a different bit of data from it.''}  [P9] and they could \textit{``Consume [the experience] in a way that [had] more meaning.''} [P7].
However, six participants commented that they had to figure out the best viewpoints by themselves, meaning that it took longer to gain the relevant information as compared to the desktop version.
We discuss this notion of reader control in Sec.~\ref{ssc:new-themes}.

\bpstart{Realism: \replaced{Realism}{Photorealism} is not important to understand the underlying data, but is still important for engagement}
In accord with Sec.~\ref{ssc:realism}, all participants agreed that more photorealistic rendering was not required to understand the underlying quantitative data, 
but that it made the experience more enjoyable and engaging.
In terms of using abstract models (e.g., cubes instead of runners), participants agreed that the feeling and perception of the data was the same, but some remarked that it became easier to make comparisons such as \textit{``[Telling] the alignment of each racer.''} [P7], mitigating similar issues identified in Sec.~\ref{ssc:sweet-spots}.
Many others pointed out flaws in these abstract models, such as them needing to \textit{``Keep in mind that it was an abstracted representation of running.''} [P7],
the inability to \textit{``Distinguish different people and separate them [as] the blocks are all the same.''} [P8],
losing the sense of emotional attachment to the phenomena as \textit{``A block is very abstract [which] you can interpret as anything coming at you, whereas a person generates some sort of emotion.''} [P9],
and that abstract looking models would be boring to look at in contrast to more realistic ones.

\bpstart{Annotation: Annotations were useful to round out the experience and were not distracting} 
In Sec.~\ref{ssc:annotation}, we raised concerns of annotations potentially being distracting. However, no participants reported any distraction or annoyance caused by them.
In fact, many complained that they were either too difficult to see or didn't provide enough information.
These participants thought annotations were important as without them \textit{``You would walk away with an unfinished idea... once you put the labels in, the information is more complete.''} [P9].
Few others thought they weren't necessary, as \textit{``VR really makes me feel the speed [and height] difference, and in that case the label doesn't really matter.''} [P5].
However, none argued for the complete removal of annotations, with all agreeing that they were at least nice to have.

\bpstart{Knowledge Transfer: The ability to apply experiences from visceralizations is uncertain, but it is still valuable to have}
\begin{wrapfigure}[10]{r}{0.285\linewidth}
\hspace{-0.325cm}
\includegraphics[trim = 20 30 0 115, width=1.1\linewidth]{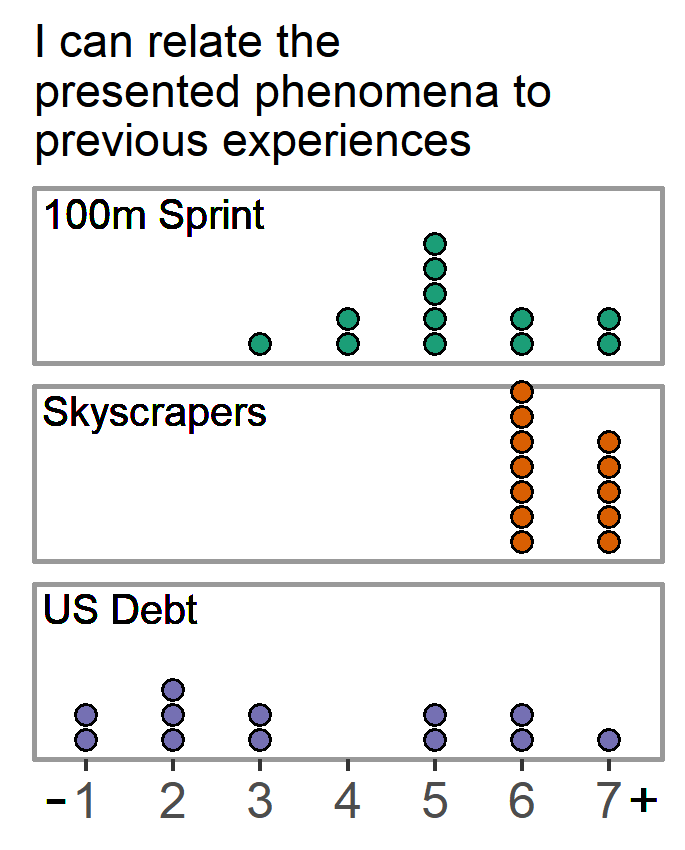}
\end{wrapfigure}
In Sec.~\ref{ssc:knowledge-transfer} we discussed the ability to transfer this deeper understanding from visceralizations into reality.
When asked if they thought they could do so, participants subjectively rated an average of 5.06 for relating to previous events (\textit{Mdn} = 6, \textit{SD} = 1.79) and 5.14 for future events (\textit{Mdn} = 5, \textit{SD} = 1.42) across all three prototypes, with \EXSix{} being the lowest rated.
Of note is that many participants who gave low or neutral scores clarified that they didn't have a prior ex-

\begin{wrapfigure}[10]{r}{0.285\linewidth}
\hspace{-0.325cm}
\includegraphics[trim = 20 30 0 10, width=1.1\linewidth]{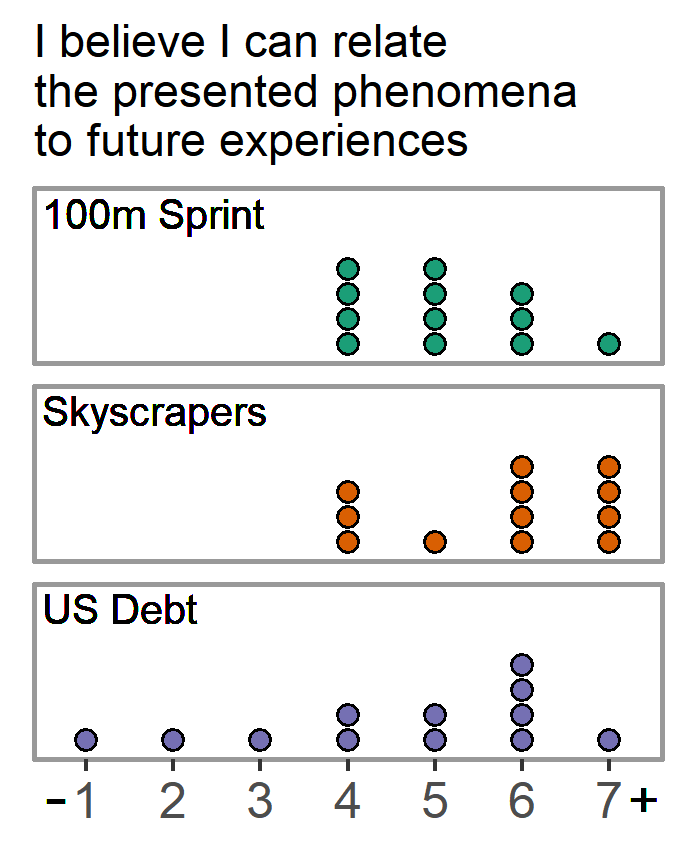}
\end{wrapfigure}
\noindent perience to compare against (e.g., people running past), or that the presented scenario was far too unlikely to ever see it in the future (e.g., stacks of \$100 bills). That said, two participants explicitly mentioned that it triggered previous memories, such as P7 having been under the actual Eiffel Tower or P8 with a specific tower in her home country.
A few were also very adamant in their ability to do so, such as \textit{``It feels I've been next to a building that's that tall, so I can compare [it to reality].''} [P2] and \textit{``That was represented really well and was really believable as compared to my [real life] experiences.''} [P7].
\replaced{O}{By and large however, o}ur measures are very subjective \added{however,} as participants were asked to hypothesize if it was possible. Most gave vague responses during the interview, but agreed that this skill was valuable to have. Regardless, we see this as a valuable means of measuring the effectiveness of data visceralization in the future (Sec.~\ref{ssc:future-work}).

\bpstart{Miniaturization: Only a few participants felt the miniaturized phenomena to be unrealistic, with others not noticing or not caring.}
\noindent 
In Sec.~\ref{ssc:mini-vs-distant} we noted differences between miniature vs distant phenomena. We asked participants if they noticed any differences in either one, making sure to be as vague as possible.
Only three participants stated that the distant skyscrapers felt like real structures which were \textit{``Easier to understand than having a miniature sculpture.''} [P9]. However this was not due to motion parallax, but due to \textit{``The representation of the hills, the elevation, and the trees [giving] me more context to the size, the scale, [and] the place.''} [P7].
All other participants gave varying responses: four saying they felt the same,
one saying that the distant felt unrealistic due to the gray background (Fig.~\ref{fig:skyscrapers}f),
and three saying they were useful simply to get an overview of the data similar to the original desktop version (Fig.~\ref{fig:teaser}c).

\bpstart{Abstract Values: Understanding exact magnitude of abstract values is difficult, but does not have to be the goal of visceralization}
In Sec.~\ref{ssc:abstract-concepts} we noted difficulties of representing abstract concepts such as money in VR, even in its physical form such as \$100 bills.
All participants agreed that it was difficult to understand the exact magnitude of \$20+ trillion. 
For some, it was the abstract nature of money being difficult to comprehend.
For others, it was the lack of quantitative information provided to them which caused them these issues. As both the trimmed desktop and our VR versions did not specify how much each stack of bills was worth, it meant that they \textit{``Just looked at towers and was told it was worth trillions of dollars, which is not a number I can really comprehend anyway.''} [P10].
Interestingly, some participants suggested solutions, such as showing incremental stages similar to the original piece \cite{Demonocracy:2017:USDebt},
or to group the \$100 bills by more familiar units such as \textit{``A block of a street, or a city.''} [P6] in a similar vein to concrete scales \cite{Chevalier:2013:UCS}.
While we did not ask if they still thought the experience was valuable, one participant commented that \textit{``It definitely gives a good perspective... it's good to know that \$20+ trillion is that much money.''} [P4].
We discuss the validity of perspective/appreciation being the goal of visceralizations in Sec.~\ref{ssc:new-themes}.

\section{Discussion} \label{sec:discussion}
In this section we discuss the most interesting and relevant themes which were raised by our participants which were not part of our critical reflection. We then discuss broader aspects of data visceralization as well as future research opportunities. 

\subsection{Insights from Participant Feedback}
\label{ssc:new-themes}
\bpstart{Viscerality is not required for data understanding but gives the experience more meaning}
When asked about feelings of viscerality, a few participants explicitly commented that it is not required for the purposes of understanding the underlying numerical measurements, as \textit{``Data and numbers are more rational, and you don't need emotions to understand rational things.''} [P11].
In contrast, many others concentrated on how visceral sensations improved their qualitative understanding of the phenomena as it made the experience feel more believable, and in some cases \textit{``Added emotion to the data... it put meaning to it.''} [P9].
When asked what types of visceral sensations they felt, common responses were the feeling of something running past you, fear of heights, \deleted{the }sense of awe/grandness, and \added{some discomfort from} flying around in the environment (particularly for \replaced{\EXThree}{skyscrapers}).
Overall, participants thought that visceral sensations improved the experience and understanding of the physical phenomena by making it \added{feel} more believable\deleted{ and giving it more meaning}, but was otherwise not necessary for learning exact values and numbers\deleted{ in the data}.

\bpstart{Data understanding and the experience go hand in hand, but numbers are not everything}
We have distinguished between the qualitative and quantitative understanding of data, with participants valuing each differently. All participants preferred VR as it provides a more immersive experience and a better qualitative understanding of what the phenomena are in reality (among other benefits).
In contrast, no participants argued for quantitative understanding (i.e., knowing the exact numerical values) being a fundamental takeaway of visceralization. Instead, it is there to further facilitate the qualitative understanding of the phenomena, as 
\textit{``The combination of both [qualitative and quantitative], they both support \replaced{each}{reach} other, one standing alone might be somewhat useful but together they're better.''} [P10].
Three participants even posited that trying to learn the exact numerical values was pointless, as 
\textit{``Who will remember the values all the time, for everything, that's crazy!?''} [P6].
In some ways, the focus on the qualitative aspects guided by quantitative values is the intended purpose of data visceralization, as we do not seek to replace conventional data visualization but instead complement it.
In this sense, an alternate way of phrasing the purpose of data visceralization is to gain a `better perspective' or `appreciation' of data, as this appreciation is oftentimes missing in data visualization.
This notion highly relates to a recent viewpoint by Wang et al. \cite{Wang:2019:ERV}, which argues for the importance of emotion when considering the value that data visualizations provide rather than only measuring analytic and knowledge-generating value. As such, the perceived value of visceralization by our participants leans into this---that it is more about the experience and the emotions it generates rather than just the numbers.

\bpstart{A balance needs to be struck between letting the user explore and guiding the user}
As described earlier, the ability to control the viewpoint was \replaced{was useful to gain}{a key benefit of visceralization, as one could attain} more nuanced insights \replaced{by having}{and have} more agency to explore. However, \replaced{more time is required to determine appropriate viewing angles in VR compared to the fixed angles on the desktop}{it naturally takes time to determine the best positions in which to view from, with participants spending much more time in VR than on the desktop (Sec.~\ref{ssc:general-results})}\replaced{, and}{. In addition,} the lack of guidance may cause users to miss important information.
Some participants argued for the use of predefined viewpoints \replaced{tied to}{associated with} specific insights as a way to alleviate this, \textit{``If you have some charts aligned to that, I can click of them and say this is the view I was looking for, so this gives me more understanding [of where] I should be going.''} [P12].
This notion is similar to author-driven and reader-driven stories \cite{Segel:2010:NVT} in determining how much guidance to provide to the user. We can reasonably say that a middle-ground would be best suited in order to retain much of the user agency that defines VR, but different considerations may be needed when used in a more storytelling context such as in immersive journalism \cite{delaPena:2010:IJI}.

\bpstart{While making comparisons were common, there are opportunities to highlight individual characteristics of standalone phenomena}
We noticed that participants almost always made comparisons between objects in the scene, such as comparing the speeds of different runners or the height of the Statue of Liberty against a stack of \$100 bills.
Three participants explicitly stated that making these comparisons was a core part of their experience, but gave mixed responses when asked if this would still be the case when only a single object/phenomena was shown. One said that it would only have the same impact \textit{``If you've had prior experience of being on heights [or] doing a sprint.''} [P9],
another that they can still gain insights by \textit{``[Going] to the top or close to the base, or see the building from the bottom of it.''} [P6].
More interestingly, P12 suggested that \textit{``If it gives me enough options to explore, like climb the wall, climb the stairs, feel different textures... the scenario is not attached to the height scenario [any more].''} In this circumstance, the data visceralization presents information beyond just the quantitative measures, but to include other characteristics of the chosen phenomena as well. While the hope is that visceralization can assist with understanding individual phenomena without the need for comparison, it is apparent there are alternative options for communicating more information should the opportunity present itself.

\subsection{\deleted{Importance of }One-to-\replaced{O}{o}ne Mapping \replaced{from}{Between} Data \replaced{to}{and} Visceralization}
Based on our critical reflection (Sec.~\ref{ssc:data-transformation}) and external feedback sessions (Sec.~\ref{ssc:user-study-findings}), it is clear that data visceralizations involving a one-to-one translation from their `ground truth' are ideal, as they most accurately portray the underlying data.
When a transformation is applied to the data, this connection to the ground truth is broken. However, it is still possible to understand comparisons between the transformed phenomena.
Moreover, many participants reported that they were still able to gain an appreciation and perspective of the data---both specifically for \EXSix{} and for data visceralizations as a whole. In this sense, despite not being the original intention, data visceralization can still provide value in instances where one-to-one mappings are not possible, either by rescaling the data or using concrete scales \cite{Chevalier:2013:UCS}.
This transformation needs to be done with care, however, as its misuse may result in misleading or deceptive experiences. For example, a visceralization which subtly re-scales data to exaggerate a certain \replaced{effect}{affect} may be taken at face value to be true. There has been extensive discussion of data visualizations potentially being deceptive, and special care needs to likewise be taken since data visceralization is intended to give an intuitive understanding of the data that is not misleading.


\subsection{Visceral vs. Emotional Experiences}
As reported by our participants, visceral sensations were tied to some emotional reaction, such as fear when looking down from a great height, or the feeling of awe when looking up at a large structure.
For the purposes of data visceralization however, it is important to differentiate between emotions stemming from visceral sensations and those from the storytelling context. That is, visceral sensations are part of the pre-attentive sensory process when experiencing the presented phenomena, in contrast to strong emotional reactions that are evoked through the narrative such as sadness or anger (e.g., \cite{Halloran:2016:FallenWW2, Lopez:2019:MassShootings}).
By treating these separately, we focus primarily on sensory fidelity and instead let the storytelling context convey any other desired emotion.



\subsection{Application Domains}
Since effective data visceralization may restrict transformation of data, there are certain domains and experiences that are particularly well suited \replaced{for}{to data} visceralization. We have already identified and demonstrated two prototypes using sports visualization, and this space could be explored more extensively. Scale comparisons of other human endeavors like architecture, engineering, and design are all enhanced by having \added{a} deeper understanding of data. In particular, understanding the relative sizes of objects while browsing online services has been often suggested as a compelling use of immersive technology. Finally, there has been captivating \deleted{and engaging }video examples (e.g., \cite{WeatherChannel:2018:Flooding}\added{)} of the impact of weather events\deleted{,} like flooding, where a one-to-one demonstration of the height of the water, or the speed of windows being blown open, could help viewers more deeply understand what happens during the event. 


\subsection{Limitations of Data Visceralization in VR}
As our presented design probes were conducted in VR, any limitations of VR directly affect the effectiveness of data visceralization. One issue in particular is that egocentric distance estimations are compressed in VR \cite{Loomis:2003:VPE, Renner:2013:PED}, possibly resulting in perceptually misleading experiences. To combat this, the use of a virtual self-avatar has been shown to improve near distance estimation \cite{Ebrahimi:2018:IEA, Mohler:2010:TEV}. It has also been shown that the size of one's virtual body influences perceived world size \cite{Hoort:2011:BBT}, and the size of one's virtual hand influences perceived size of nearby objects \cite{Linkenauger:2013:WWT} which can in turn help improve size estimation \cite{Jung:2018:OMH}. As visceralizations are reliant on the accurate perception of virtual objects, these techniques and others similar in the VR literature can be used.
\added{Another limitation is the trade-off between visual fidelity, dataset size (if applicable), and performance. This was notably of concern in \EXFive, as rendering thousands of objects with complex meshes resulted in poor performance. While we refer to general VR development guidelines to overcome these challenges (e.g., \cite{Facebook:PerformanceGuidelines}), we note that graphical quality may ultimately not be important for data visceralization (Sec.~\ref{ssc:general-results}).}

\subsection{Future Work} \label{ssc:future-work}
\bpstart{Evaluation}
\deleted{Our work involved eliciting qualitative feedback from participants. }While responses \added{from our participants} supported the notion that data visceralizations were engaging and impactful, \replaced{their feedback was inherently}{any feedback we received was inherently} subjective. \replaced{We }{Our work} discussed several experimental measures and factors which may be considered \replaced{in future studies}{when evaluating data visceralizations in the future}. These include: users' capability to perceive and comprehend the qualitative `ground-truth', time until this comprehension is reached, the transferability and generalization of learnings into the real-world, and the effects of numerous factors (e.g., transformation from data to visceralization, number and types of annotations, level of realism) on understanding. By more formally considering and measuring these in future user studies can we more thoroughly evaluate the effectiveness and appropriate use of data visceralization\added{---particularly in comparison to similar techniques such as those in Sec.~\ref{sec:related-work}}.

\bpstart{Data visceralization beyond VR}
Advancements \replaced{in}{and commoditization of} immersive technologies opens many new opportunities for data visceralization. For example, researchers have explored how non-visual senses can be stimulated to convey data, such as olfaction (e.g., \cite{Patnaik:2019:IOH, Washburn:2004:ODI}), gustation (e.g., \cite{Khot:2015:TDP}), and haptic simulation (e.g., \cite{Batter:1972:GCD, Brooks:1990:PGD:97880.97899}).
\replaced{AR also}{In addition, AR} offers unique opportunities and challenges, as the real-world acts as an anchor for all measurements and scales. While this may improve relatability of the data as the visceralization is in the context of the user's environment, it may restrict the scope of phenomena which can be represented---particularly for large scale objects such as skyscrapers.

\bpstart{Combining narrative storytelling and data visceralization}
As was the original intention and reaffirmed by participants, data visceralizations can fill in the `ground truth' understanding \deleted{of the data }oftentimes missing \replaced{in}{from} data visualization. As such, it is important to determine how \deleted{best }to \replaced{combine the two}{integrate these together} without sacrificing \deleted{much of }the accessibility and convenience of web-based data stories. As described in Sec.~\ref{sec:related-work}, recent work has begun exploring the use of immersive experiences \replaced{for data storytelling}{to tell stories with data} \cite{Bastiras:2017:CVR, Isenberg:2018:IVD, Ren:2018:XIC}, with some existing stories already doing so using mobile-based AR (e.g., \cite{NewYorkTimes:2018:AugmentedRealityOlympics}). Data visceralization can both contribute to and utilize these immersive data-driven storytelling techniques in the future.

\bpstart{Towards a general data visceralization pipeline}
\replaced{D}{At present, d}evelopment of our VR prototypes required skills in design, 3D modeling, and programming in specialized environments such as Unity3D.
\added{That said, many pre-made assets were used to hasten development, resulting in development times of around a week for each prototype (excluding reflection/evaluation).}
\replaced{During this process}{During our critical reflection}, we fell into similar development patterns which we portray in Fig.~\ref{fig:pipeline}.
Sitting in parallel to the visualization pipeline \cite{Card:1999:RIV}, we pass data into the data visceralization pipeline and directly map it onto attributes of objects within the virtual environment. This \deleted{process of }simulation may use animation to \replaced{encode}{include other} information such as \replaced{speed in \EXOne{}}{time, as seen in \EXOne{} in order to portray speed}.
After adding appropriate background imagery to \deleted{better} contextualize the simulated phenomena, we then render these at the desired level of realism.
Finally, we annotate the experience with the original quantitative data to provide perspective and context to the experience.
The output is then experienced on \replaced{a}{an appropriate} VR \deleted{(or AR)} \replaced{device}{display}---ideally when complementing an existing data visualization or story. While nothing is currently automated in this pipeline, the structure may help teams organize where different contributions can occur.
\section{Conclusion} \label{sec:conclusion}
We introduce and define the concept of data visceralizations as a complement to data visualization and storytelling, in order to facilitate an intuitive understanding of the underlying `ground-truth’ of data.
We explore the concept through the iterative design and creation of six design probes, which we critically reflect on as authors and gather external feedback and opinions from participants. Through this, we identify situations where data visceralization is effective, how transforming the data or representing abstract data may be problematic for visceral experiences, and considerations of many factors for data visceralization as a whole. We also identify future opportunities, such as formally evaluating data visceralization in a user study, and how future technologies can extend the concept. We hope that this work will spawn both future experiences for understanding data as well as deeper investigation into how to best create and more formally evaluate these experiences.



\bibliographystyle{abbrv-doi}

\bibliography{bibliography}
\end{document}